\documentclass[a4paper,12pt]{article}

	\usepackage[utf8]{inputenc}
	\usepackage[dvipsnames]{xcolor}
	\usepackage{booktabs}
	\usepackage{graphicx}
	\usepackage{natbib}
	\usepackage{verbatim}
	\usepackage{longtable}
	\usepackage{tabularx}
	\usepackage{multicol}
	\usepackage[english]{babel}
	\usepackage{array}
	\usepackage{graphicx}
	\usepackage{caption}
	\usepackage{subcaption}
	\usepackage{float}
	\usepackage{tikz}
	\usepackage[font=itshape]{quoting}
	\usepackage{a4}
	\usepackage{fancyhdr}
	\usepackage{fontenc}
	\usepackage{hyperref}
	\usepackage{amssymb,amsfonts}
	\usepackage{lscape}
	\usepackage{amsmath}
	\usepackage{mathtools}
	\usepackage{amsthm}
	\usepackage{threeparttable}
	\usepackage{rotating}
	\usepackage{longtable}
	\usepackage{epstopdf}
	\usepackage{enumerate}
	\usepackage[page,title]{appendix}
	\usepackage{fullpage}
	\usepackage{algorithm}
	\usepackage[noend]{algpseudocode}
	\usepackage[shortlabels]{enumitem}
    \usepackage{dsfont}

	\newtheorem{theorem}{Theorem}
	\newtheorem{lemma}{Lemma}[section]
	\newtheorem{assumption}{Assumption}
	\newtheorem{corollary}{Corollary}

	\newtheorem{proposition}{Proposition}
	
	\theoremstyle{definition}
	\newtheorem{definition}{Definition}

	\newtheorem{remark}{Remark}

	\def\X{\mathcal{X}}
	\def\Y{\mathcal{Y}}

	\def\R{\mathbb{R}}
	
	\def\NN{\mathcal{N}}

	\def\d{\mathrm{d}}

	\renewcommand{\epsilon}{\varepsilon}
	\def\sigmatil{\tilde{\sigma}}
	\def\lambdatil{\tilde{\lambda}}
	\def\Btil{\tilde{B}}

  \graphicspath{{Figures}}

	\title{On spatial systems of cities}
    
\date{}
	\author{Gianandrea Lanzara\thanks{University of Bologna, \texttt{g.lanzara@unibo.it}}~ and Matteo Santacesaria\thanks{MaLGa Center, Department of Mathematics, University of Genoa, \texttt{matteo.santacesaria@unige.it}}}

\begin{document}
\maketitle

\begin{abstract}
Are there multiple equilibria in the spatial economy? This paper develops a unified framework that integrates systems of cities and regional models to address this question within a general geographic space. A key feature is the endogenous formation of commuting areas linking a continuum of residential locations to a finite set of potential business districts. Using tools from computational geometry and shape optimization, we derive sufficient conditions for the existence and uniqueness of spatial equilibria. For plausible parameter values, urban location is indeterminate, but, conditional on an urban system, city sizes are uniquely determined. The framework reconciles seemingly conflicting empirical findings on the role of geography and scale economies in shaping the spatial economy.

\end{abstract}

\vspace{1.5cm}

\noindent
JEL-Classification: F10, R12, R13, C02
\vspace{1.5cm}

\newpage

\section{Introduction}

A fundamental question in economic geography is whether the spatial distribution of economic activity is uniquely determined by geographic characteristics, or whether increasing returns to scale are strong enough to generate multiple equilibria. This question is critical for understanding the roots of spatial inequalities, how local economies respond to temporary or permanent shocks, and the scope for policy interventions. Yet, the answer remains unclear, as both theoretical and empirical studies point in different directions \citep{lin2022}. 

Multiple equilibria are a central theme of the ``new economic geography'' literature, both in the regional models of \cite{krugman1991} and \cite{helpman1998} and systems of cities models such as \cite{fujita1997}.\footnote{\cite{krugman1993} studies the location of a single urban center on a segment or a disk and finds that there is a range of equilibrium urban locations. \cite{fujita1995} take the location of the metropolis as given, and investigate under what conditions the monocentric equilibrium is sustainable. \cite{fujita1997} extend this approach to a setting with multiple urban centers and population growth. See \cite{fujita2002} for a comprehensive review of the new economic geography literature.} By contrast, following \cite{allen2014} and \cite{redding2016}, a large literature in quantitative economic geography has mainly focused on environments where the equilibrium is unique.\footnote{In a broad class of quantitative spatial models, a sufficient condition guarantees uniqueness, and this condition is typically satisfied in empirical settings \citep{allen2024}. See \cite{allen2025} for a recent review of the literature on quantitative regional models.} However, while the equilibrium properties of regional models are well-understood, we lack comparable analytical results for spatial systems of cities models.

Searching for multiple equilibria in the data has also led to conflicting conclusions. On the one hand, the findings in \cite{davis2002} support the idea that geographic characteristics determine the spatial structure of the economy; on the other hand, \cite{bleakely2012} and \cite{michaels2018} present evidence in favor of a world where increasing returns to scale play a dominant role.  

In this paper, we propose a model of a spatial urban system that integrates systems of cities and regional models in a setting with realistic geography and provide a novel characterization of its equilibrium properties. Based on our analytical results, we offer a new perspective that helps reconcile previous views on the existence of multiple equilibria in the spatial economy.

A key feature of the model is a distinction between a \textit{continuum} of residential locations and a \textit{finite} set of potential business districts. Agents with identical preferences optimally choose a residential location and a commuting destination where they supply labor and purchase final goods, subject to commuting costs. As a result, the partition of space into commuting areas and the set of business districts that successfully attract commuters are equilibrium outcomes.\footnote{In this respect, the model differs from quantitative urban models with heterogeneous individual valuations for residence-workplace pairs, such as \cite{ahlfeldt2015} and the related literature. Because of the extreme-value assumption on individual shocks, in these models each production location attracts commuters from all residential locations. See \cite{redding2025} for a recent review. } Apart from this feature, the other components of the model are standard: it incorporates the gravity structure of the Armington trade model, free labor mobility, and heterogeneous values of residential amenities and labor productivity across locations, both of which may be subject to local externalities. Yet, the characterization of the equilibrium properties is challenging because the boundaries between commuting areas trace piece-wise smooth curves in the plane, and the shape of these areas is determined in general equilibrium. We apply tools from computational geometry and shape optimization to provide novel sufficient conditions for the existence and uniqueness of equilibria in a spatial systems of cities model. 

A key insight of our analysis is that it is important to distinguish between two sources of multiple equilibria. Firstly, there may exist multiple equilibria in the spatial structure of the urban system, i.e., in the location of sites hosting economic activity. This source of multiplicity is governed by a trade-off between the strength of scale economies in production and consumers' taste for variety. If increasing returns to scale exceed a certain cutoff, economic activity clusters into a subset of potential sites, making the location of urban centers essentially indeterminate. Secondly, for a given urban system, there may be multiple equilibrium distributions of labor across its active business districts. As in regional models, it is possible to derive a sufficient condition for uniqueness that is satisfied when dispersion forces, including the strength of residential spillovers, prevail over agglomeration forces. 

Thus, there is a region of the parameter space in which the model admits multiple equilibria in the spatial arrangement of business districts that host economic activity and a unique equilibrium in the distribution of labor across these active sites. This scenario deserves attention for two reasons. First, the corresponding region overlaps with the range of estimates available in the literature, provided that commuting costs are large. Specifically, increasing returns to scale must be sufficiently strong for firms to agglomerate, yet not so strong as to overwhelm dispersion forces when it comes to the spatial allocation of labor. Second, it aligns with several empirical regularities that are often viewed as providing conflicting evidence regarding the existence of multiple equilibria and the roles of geography versus increasing returns to scale. 

To illustrate, in this scenario the model predicts that temporary changes to the city-size distribution have only transient effects of the economy, provided that they do not completely wipe out urban centers, consistent with the evidence from \cite{davis2002} on the recovery of Japanese cities after World War II bombings.\footnote{\cite{davis2002} find that, despite the extensive destruction, Japanese cities returned to their long-run relative city sizes about fifteen years after World War II. \cite{brakman2004} obtain similar results for German cities. In the same context, but using different methods, \cite{bosker2007} instead find that World War II bombings had permanent effects on the German city-size distribution.} Conversely, historical episodes that cause the birth or death of urban centers may permanently lead to the emergence of a new equilibrium urban system, in line with \cite{bleakely2012}'s finding that cities formed at portage sites in the U.S. Southeast and Midwest persist as population centers today, despite the decline of river transport. These predictions are also consistent with \cite{michaels2018}'s study of  British and French urban systems after the fall of the Western Roman Empire. In Britain, where urbanization collapsed for over a century, the urban network reconfigured around locations with better coastal access, in line with the increased importance of water transport. In France, where urban life experienced greater continuity, medieval towns are more likely to occupy the same sites as their Roman predecessors, but cities with better coastal access gained population. By yielding multiple equilibria in urban \textit{location} and a unique equilibrium in urban \textit{size}, our framework provides a unified explanation for these patterns.

The spatial setup of this paper differs from the previous literature, that has either considered continuous spaces or finite locations. An exception within the new economic geography tradition is \cite{takayama2020}, who study the emergence of an urban system on a seamless circular economy with a continuum of residential locations and a finite set of equidistant potential urban sites. In the more recent quantitative literature, two closely related papers are \cite{nagy2020} and \cite{nagy2023}. In \cite{nagy2020}, the spatial economy consists of a finite number of locations where people live and produce a differentiated good, while trading activities take place at a fixed subset of these locations serving as trading places. In \cite{nagy2023}, a continuum of locations specialize in either farm or nonfarm activities, with rural hinterlands forming around nonfarm centers that also serve as trading hubs. To the best of our knowledge, we are the first to provide existence and uniqueness conditions for a spatial system of cities with endogenous commuting areas, either on simplified or general geographies. 

By using additively weighted Voronoi diagrams to describe the partition of space into commuting areas, our work is also related to a small number of papers using tools from computational geography in spatial economics -- see, for instance, \cite{allen2023}, \cite{lanzara2023}, and \cite{oberfield2024}. In particular, \cite{allen2023} uses centroidal Voronoi tessellations to study the evolution of national boundaries and capital cities in a spatial setting with heterogeneous locations. In this model, a political economy mechanism determines the assignment of locations to capital cities and the location of capital cities within their polities. By contrast, in our model the location of cities results from the endogenous selection of a subset of potential sites. In \cite{lanzara2023}, we applied Voronoi diagrams to a setting with costless trade across urban sites.

More generally, our analysis bridges earlier theoretical traditions in urban and economic geography. Firstly, relative to the systems of cities framework of \cite{henderson1974}, the set of monocentric economies is ordered in space and connected through costly trade and adjacent commuting areas. In turn, we do not incorporate multiple sectors in the production of urban goods, thus ruling out the possibility of different ``types'' of cities. Secondly, relative to new economic geography models, such as \cite{fujita1995} and \cite{fujita1997}, our framework accommodates more general geographies and parameter spaces, nesting the scenario with monopolistic competition and fixed costs of production as a special case; like these studies, we can easily introduce an agricultural sector as the source of congestion in our framework. Finally, if we fix the urban system \textit{a priori}, the model resembles a regional framework, except that the endogenous boundaries between spatial units act as additional agglomeration force. When either commuting costs are large or business districts are far apart, so that border effects wane out, this agglomeration force vanishes and the model behaves like a regional framework with discrete locations, as in \cite{helpman1998} and \cite{allen2014}.

\section{A spatial model of urban systems}\label{sec:model}

\subsection{Setup}
Let $\X$ be an open, bounded, connected subset of the Euclidean plane $\R^2$ and let $\Y \subset \X$ be a finite set of $n \geq 2$ points. There is a measure $L > 0$ of ex ante identical agents, each endowed with one unit of labor. These agents reside in locations in $\X$ and commute to a location in $\Y$ for labor supply and consumption.\footnote{In Appendix \ref{app:homeconsumption}, we carry out the analysis under the assumption that agents consume the differentiated varieties at their residential location, thus incurring an additional transport cost.} Each $y \in \Y$ is endowed with the technology to produce a distinct variety of a final good. We refer to the elements of $\X$ as residential locations and to the elements of $\Y$ as business districts.

Locations are heterogeneous in terms of amenities and labor productivity. First, we define the measure $b\colon \X \to \R_{+}$ to be the value of the residential amenities enjoyed at residential locations  in $\X$. Second, we define $A = \{A_i\}_{i = 1}^n$, $A \in \R_{+}$, to be the vector of labor productivities at business districts in $\Y$.

Finally, locations differ in terms of the cost of commuting to and trading with other locations. These costs take the Samuelson iceberg form. Specifically, commuting costs are described by a function $D\colon \X \times \Y \to [1, \infty)$, such that an agent living at $x$ and commuting to $y_i$ is able to enjoy fraction $1/D(x,y_i)$ of the local residential amenities.\footnote{Alternatively, we could assume that commuting reduces the efficiency units of supplied labor. This assumption would not affect our analysis.} Also, trade costs are described by a function  $T\colon \Y \times \Y \to [1, \infty)$, such that each unit of the good shipped from $y_i$ results in $1/T_{ij}$ units reaching their destination at $y_j$. 

Agents share identical preferences defined over the consumption of the differentiated varieties of the good. They order consumption baskets $c = \{c^j\}_{j = 1}^{n}$ according to the utility function $U\colon \R_{+}^{n} \to \R_{+}$, $U(c) = ( \sum_{j=1}^{n} (c^j)^{\frac{\sigma - 1}{\sigma }} )^{\frac{\sigma}{\sigma - 1}}
$, with constant elasticity of substitution $\sigma > 1$ between varieties. A continuum of competitive firms produce these varieties with constant returns to scale and using labor as the only factor of production. For each $y_i \in \Y$, the technology is such that aggregate output $Y_i$ is given by $Y_i = A_i L_i$, where $L_i$ denotes the labor demand at $y_i$. 

Agents optimally choose a residential location in $\X$ and a commuting destination in $\Y$, taking into account the indirect utility of consumption, the value of residential amenities, and commuting costs. Relocating to a different residence-workplace pair is costless. 

Because they are ex ante identical, all residents at $x$ will choose the same commuting destination. We will use $\Omega_i \subseteq \X$ to denote business district $y_i$'s commuting area. In equilibrium, some of the business districts in $\Y$ may not receive any commuters, i.e., $\Omega_i = \emptyset$. In this case, we say that $y_i$ is a vacant business district, as opposed to active business districts, where production and consumption take place. We will use $\Y^* \subseteq \Y$ to denote the set of $n^* \leq n$ active business districts. 

We first describe an equilibrium with a given subset $\Y^{*} \subseteq \Y$ of $n^* \leq n$  active business districts. Then, we provide conditions for such an equilibrium to be sustainable in face of deviations directed to vacant business districts in $\Y \setminus \Y^*$.  Without loss of generality, we order the elements of $\Y$ such that $y_i \in \Y^{*}$ for $i = 1, \dots, n^{*}$. 
	
\paragraph{Consumer's problem.} An agent solves 
\begin{equation}\notag
\max_{ \{c^j\}_{j = 1}^{n^*} } U(c) \quad \text{subject to } \sum_{j = 1}^{n^*} p^j c^j \leq w,
\end{equation}	
where the wage rate $w$ and the price vector $\{p^j\}_{j=1}^{n^*}$ are taken as given, and $c^j = 0$ for the varieties $j = n^*+1,\dots, n$ not in production.

It is convenient to define the demand functions in terms of the business district where the agents consume, rather than prices and wages. Let $p^j_i$ denote the unit price of a variety produced at $y_j  \in \Y^*$ and delivered at $y_i \in \Y^*$. For each $y_j$, the consumer's problem yields a demand function $C^j\colon \Y^* \to \R_{+}$, such that 
\begin{equation}\label{eq:C}
C^j_i = (p^j_i)^{-\sigma} P_i^{\sigma - 1} w_i
\end{equation}
gives the individual demand at $y_i$ for the variety imported from $y_j$, where
\begin{equation}\label{eq:P}
	P_i = \left( \sum_{j = 1}^{n^*} (p^j_{i})^{1-\sigma} \right)^{\frac{1}{1-\sigma}}
	\end{equation}
denotes the ideal price index of the consumption good in $y_i$. 

\paragraph{Firm's problem.} Let $p_i$ denote the factory-gate price of the variety produced at $y_i \in \Y^*$. Because firms are price-takers and deliver goods subject to iceberg trade costs,
\begin{equation}\label{eq:p}
	p_i = \frac{w_i}{A_i}, \quad p^i_j = T_{ij}p_i. 
	\end{equation}
At this price, firms earn zero profits, 
\begin{equation}\label{eq:zeroprofit}
p_i Y_i = w_i L_i.
\end{equation}	

\paragraph{Spillovers.}	
	Labor productivity and residential amenities may be subject to local externalities. 	
	Let $\{\bar{A}_i\}_{i=1}^n$ represent the exogenous component of labor productivity, such that $\bar{A}_i$ captures the fixed characteristics of a business district $y_i \in \Y$ that make the local labor supply more or less productive. We assume 
	\begin{equation}\label{eq:A}
	A_i = \bar{A}_i L_i^{\alpha},
	\end{equation} 
	where $\alpha \in \R$ parametrizes the effect of local labor supply on labor productivity. 
	
	Also, let $\bar{b}\colon X \to \R_{+}$ represent the exogenous component of residential amenities, such that $\bar{b}(x)$ captures the fixed characteristics of a residential location $x \in \X$ that make it more or less attractive for workers. We assume 
	\begin{equation}\label{eq:b}
	b(x) = \bar{b}(x)\ell(x)^{\beta},
	\end{equation} 
	where $\beta \in \R$ parametrizes the effect of local population on residential amenities. 

    In what follows, we restrict our attention to the case where $\alpha > - 1$ (total output increases with the quantity of the labor input) and $\beta < 0$ (the level of residential amenities decreases with the size of resident population).
		
\paragraph{Market clearing.} Goods and labor markets clear when supply equals demand at all business districts: 
\begin{align}
	\label{eq:gmc}
	Y_i = A_i L_i &= \sum_{j = 1}^{n^*} T_{ij} C^i_j \int_{\Omega_j} \ell(x), \quad i = 1,\dots,n^*,  \\
	\label{eq:lmc}
	L_i &= \int_{\Omega_i} \ell(x) \d x, \quad i = 1, \dots, n^*. 	
 \end{align}	

 \paragraph{Welfare.}
 The agent's overall welfare is given by  the product of her consumption utility and the level of local residential amenities, with a discount due to the cost of commuting. At the optimal consumption choice \eqref{eq:C}, we can express the welfare of an agent residing at $x \in \X$ and commuting to $y_i \in \Y^*$ with the indirect utility function $V\colon \X \times \Y \to \R_{+}$,
	\begin{equation}\label{eq:V}
	V(x, y_i) =  \frac{b(x) w_i}{D(x, y_i) P_i}.
	\end{equation}
Agents choose a residential location $x$ and a commuting destination $y_i$ in order to maximize their indirect utility. 

\paragraph{Equilibrium.} We say that a \textit{geography} is the collection $\{\bar{b}, \bar{A}, T, D\}$ of the exogenous attributes of the space $\X$. We also say that a \textit{commuting pattern} is a collection of commuting areas $\{\Omega_i\}_{y_i \in \Y}$ and a density $\ell$ of residents. Clearly, a commuting pattern immediately gives the labor supply $\int_{\Omega_i} \ell(x) \d x$ to each $y_i \in \Y$. 

For known geography and parameters, and a given commuting pattern, a \textit{market equilibrium} consists of consumption, output, labor demand and labor supply levels, wages, and prices such that: (i) agents make optimal consumption choices (see \eqref{eq:C} and \eqref{eq:P}) (ii) firms make zero profits (see \eqref{eq:p} and \eqref{eq:zeroprofit}); (iii) amenities and productivity are taken as given, although they may be subject to local externalities (see \eqref{eq:A} and \eqref{eq:b}); (iv) and, finally, goods and labor markets clear (see \eqref{eq:gmc} and \eqref{eq:lmc}).

In the analysis below, we consider two notions of spatial equilibrium. First, we \textit{posit} an economy with a set $\Y^* \subseteq \Y$ of active business districts, and  we restrict the choice of a commuting destination to business districts in $\Y^*$. That is, agents solve
\begin{equation}\label{eq:maxV}
\max_{x \in \X, y_i \in \Y^*} V(x, y_i).
\end{equation} 
 A \textit{$\Y^*$-centric equilibrium} consists of a market equilibrium and a commuting pattern that satisfies \eqref{eq:V}, \eqref{eq:maxV}, and the aggregate population constraint 
\begin{equation}\label{eq:aggpop}
	\sum_{y_i \in \Y^{*}} \int_{\Omega_i} \ell(x) \d x = L.
\end{equation}
This equilibrium concept neglects the possibility that agents may find it optimal to relocate to vacant business districts. In the next section, we develop an appropriate notion of (potential) welfare at vacant business districts to handle this possibility. A \textit{spatial equilibrium} is a $\Y^*$-centric equilibrium that is also consistent with agents solving \begin{equation}\label{eq:maxV2}
\max_{x \in \X, y_i \in \Y} V(x, y_i),
\end{equation} 
 where the choice set of commuting destinations now encompasses all sites in $\Y$. 
 
\paragraph{Restrictions on the geography.} Before we proceed, we impose a number of assumptions on the geographic component of the model. 
 \begin{assumption}\label{ass:Ab} Labor productivity and residential amenities are bounded, trade costs are symmetric and bounded. That is,
\begin{enumerate}[label={~\alph*)},
  ref={\theassumption.\Alph*}]
\item there exist $\bar{b}_{min}$ and $\bar{b}_{max}$ such that $0 < \bar{b}_{min} \leq \bar{b}(x) \leq \bar{b}_{max}$, all $x \in \X $; \label{ass:b}
\item there exist $\bar{A}_{min}$ and $\bar{A}_{max}$ such that $ 0 < \bar{A}_{min} \leq \bar{A}_i \leq \bar{A}_{max}$, all $y_i \in \Y $; \label{ass:A}
\item $T_{ij} = T_{ji}$, and there exists $T_{max}$ such that $1 \leq T_{ij} \leq T_{max}$ for all $y_i, y_j \in \Y$. \label{ass:T}
\end{enumerate}
\end{assumption}	
For each $y_i \in \Y$, let $d_i\colon \R^2 \to \R_+$ be a continuous function that assigns to a point $x \in \R^2$ a non-negative value $d_i(x)$, representing the ``distance" from residential location $x$ to business district $y_i$. We assume a specific functional form for the cost of commuting.  
\begin{assumption}\label{ass:D}
Commuting costs are an exponential function of distance, i.e.:
\begin{equation}\notag
D(x, y_i ) = e^{\delta d_i(x)} \quad \text{all } x \in \X \text{ and } y_i \in \Y, \quad \delta > 0.
\end{equation}
\end{assumption}
Following \cite{geiss2013} we require the system of distance functions $\{d_i(\cdot)\}_{y_i \in \Y}$ to be \textit{admissible} (see \cite{geiss2013}, Definition 1). In essence, this condition states that as $c \in \R$ increases with $y_i \neq y_j$, the Lebesgue measure of the set $\{x \in \X: d_i(x) - d_j(x)  < c\}$ increases continuously (we provide the technical details in Appendix \ref{app:voronoi}). We introduce this requirement, along with two additional conditions, in the next assumption, with $B_r(y_i)$ denoting a ball centered at $y_i$ with a radius of $r$ in Euclidean distance.
\begin{assumption}\label{ass:d} The system of distance functions $\{d_i\}_{y_i\in \Y}$ has the following properties:
\begin{enumerate}[label={~\alph*:},
  ref={\theassumption.\alph*}]
  \item \label{ass:admiss} it is admissible \citep{geiss2013}
  \item \label{ass:trineq} $d_i(x) \leq d_i(y_j) + d_j(x)$ for all $y_i, y_j \in \Y$ an $x \in \X$ (triangle inequality) 
\item \label{ass:doublelip} there exists $r > 0$ such that for each $y_i \in \Y$, 
$c\|x-y_i\| \leq d_i(x) \leq C \|x-y_i\|
$
holds for all $x \in B_r(y_i)$, with constants $c, C >0$ possibly depending on $r$.
\end{enumerate}
\end{assumption}
Assumptions \ref{ass:admiss} guarantees that the set of commuters who are indifferent between any two commuting destinations has measure zero (see Lemma 1 in \cite{geiss2013}, also reported in Appendix \ref{app:voronoi}). Assumption \ref{ass:trineq} is a version of the triangle inequality, and simplifies the condition for a business district to have a positive commuting area. Finally, Assumption \ref{ass:doublelip} says that we can control $d_i(\cdot)$ with the Euclidean distance in a neighborhood of $y_i$. We will use it to study the limit behavior of the model as commuting area $\Omega_i$ becomes arbitrarily small for some business district $y_i$ in $\Y^*$.

Finally, for some of our results we will also impose the following counterpart of Assumption \ref{ass:D} for trade costs: 
\begin{assumption}\label{ass:T2}
Trade costs are an exponential function of distance, i.e. there exists $\tau>0$ such that
\begin{equation}\notag
T_{ij} = e^{\tau d_j(y_i)} \quad \text{all } y_i, y_j \in \Y.
\end{equation}
\end{assumption}
Combined with Assumption \ref{ass:T}, this assumption implies that the distance functions are symmetric between the elements of $\Y$, i.e., $d_i(y_j) = d_j(y_i)$ for all $y_i, y_j \in \Y$.

\paragraph{Endogenous commuting patterns.} The main technical challenge in the analysis of the model is to characterize the endogenous shape of the commuting areas, whose boundaries trace piece-wise smooth curves in the plane, and their general equilibrium interactions with the other economic variables.\footnote{For a given commuting pattern, the model becomes a trade model with heterogeneous locations and bilateral trade costs, whose equilibrium properties are well-understood \citep{allen2020}. In particular, given the previous assumptions, it is possible to reduce the conditions for a market equilibrium into a system of equations that uniquely determines the equilibrium prices. } We now derive some implications of the commuting problem \eqref{eq:maxV} that allow us to overcome these difficulties.

First, consider the choice of a commuting destination for a given density of residents $\ell$. Under Assumption \ref{ass:D}, and using equation \eqref{eq:V}, we have that	\begin{align*}
	V(x, y_i) & > V(x, y_j) \iff\\
	 d_i(x) - \frac{1}{\delta} \log \frac{w_i}{P_i} & < d_j(x) - \frac{1}{\delta} \log \frac{w_j}{P_j}, \qquad \text{all } y_i, y_j \in \Y^*, j \neq i.
	\end{align*}
These inequalities define an \textit{additively weighted Voronoi tessellation} over $\X$, where the vector of Voronoi weights $\lambda \in \R^{n^*}$ is given by
\begin{equation}\label{eq:lambda}
	\lambda_i = \frac{1}{\delta} \log \frac{w_i}{P_i},  \qquad i = 1,\dots, n^*.
	\end{equation}
For any $\lambda \in \R^{n^*}$, we can construct $\Omega_i \subseteq \X$ consistent with \eqref{eq:maxV}, as		\begin{equation}\label{eq:Omega}
	\Omega_i(\lambda) = \left\{ x \in \X:  d_i(x) - \lambda_i < \min_{\substack{j=1\dots {n^*} \\ j \neq i}} \{ d_j(x) - \lambda_j \} \right\}.
	\end{equation} 
Three remarks are useful about this construction. First, commuting areas are defined by strict inequalities, whereas the equation  $V(x, y_i) = V(x, y_j)$ defines the set of locations where residents are indifferent between commuting to $y_i$ or $y_j$. Under Assumption \ref{ass:admiss}, these sets have measure zero and are irrelevant for equilibrium outcomes (see Appendix \ref{app:voronoi}). Second, adding the same constant to all Voronoi weights leaves the resulting tessellation unchanged. 
Third, and finally, certain weight vectors in $\R^{n^*}$ will result in a subset of business districts failing to attract any commuters.  

Next, consider the choice of a residential location for a given vector of Voronoi weights $\lambda$. With $\beta < 0$, the choice over $\X$ implies that all residential locations host a positive measure of agents and yield the same level of welfare at equilibrium, that is, $V(x, y_i) = V(x^{\prime}, y_j) = V$ for all $x, x^{\prime} \in \mathcal{X}$ and $y_i, y_j \in \Y^{*}$, such that $x \in \Omega_i$ and $x^{\prime} \in \Omega_j$, where the scalar $V>0$ is the common level of welfare in the economy.\footnote{Intuitively, given equations \eqref{eq:b} and \eqref{eq:V}, the indirect utility in $x$ goes to infinity as the resident population in $x$ approaches zero. } We define
\begin{equation*}\label{eq:B}
	B_i(\lambda) = \left[ \int_{\Omega_i(\lambda)} \left( \frac{ \bar{b}(x)}{D(x, y_i)}\right)^{-\frac{1}{\beta}}\d x \right]^{-\beta}, \; i = 1, \dots, n^*.
	\end{equation*}
as the aggregate residential amenities offered by commuting area $\Omega_i$. 
Using equations \eqref{eq:b}, \eqref{eq:lmc}, and \eqref{eq:V}, we can write the equalized level of welfare within $\Omega_i$, $V_i$, as
\begin{equation}\label{eq:Vi}
	V_i = B_i(\lambda) \frac{w_i}{P_i} L_i(\lambda)^{\beta},
\end{equation}
with $L_i(\lambda) = \int_{\Omega_i(\lambda)} \ell(x) \d x $, and the welfare equalization across commuting areas as
\begin{equation}\label{eq:weq}
    V_i = V, \quad i, \dots, n^*.
\end{equation}
These expressions take the same form as in a regional model with discrete locations, except that the level of residential amenities is endogenous because it varies with the commuting tessellation. Because they are related to real wages, the Voronoi weights encode all relevant information about the equilibrium assignment of residents to business districts, as well as the equilibrium allocation of aggregate labor supply across business districts. We leverage this insight to characterize the spatial equilibria of the model.

\section{Existence and uniqueness of spatial equilibria} 

In this section, we derive our main analytical results. First, we show that we can represent the equilibria of the model in terms of a set of equations that only depend on the vector of Voronoi weights. Second, we provide sufficient conditions for the existence and uniqueness of $\Y^*$-centric equilibria in terms of the strength of agglomeration and congestion externalities (respectively, $\alpha$ and $\beta$), the elasticity of substitution ($\sigma$) and the size of commuting costs ($\delta$). Third, we provide conditions for a $\Y^*$-centric equilibrium to sustain a spatial equilibrium. The following composite parameters will play a key role in the analysis:
\begin{equation*}
    \begin{aligned}
        \gamma_1 &= 1 - (\sigma - 1) \alpha - \sigma \beta, \quad 
        \gamma_2 = 1 + \sigma \alpha + (\sigma - 1) \beta, \quad 
        \sigmatil = \frac{\sigma - 1}{2 \sigma - 1}, \\ 
        \phi_1 &=\frac{(1-(\sigma-1)\alpha )}{\beta}, \quad 
        \phi_2 =  - \frac{(1 + \sigma\alpha ) }{\beta}.
    \end{aligned}
\end{equation*}

\subsection{Using the Voronoi weights to describe the equilibria}

Our first result focuses on $\Y^*$-centric equilibria.
\begin{proposition}\label{prop:lambda_eq}
Suppose that Assumptions \ref{ass:Ab}-\ref{ass:admiss} hold, and fix $\Y^* \subseteq \Y$. Then, the vector of Voronoi weights $\lambda$ and the level of welfare $V$ that are consistent with a $\Y^*$-centric equilibrium coincide with the solutions of the system of equations
\begin{align}\label{eq:lambda_eq}
	e^{-\frac{\delta}{\beta} \tilde{\sigma}\gamma_1 \lambda_i } &= V^{(\sigma-1){\frac{\alpha}{\beta}}} \sum_{y_j \in \Y^*} T_{ij}^{1-\sigma} \bar{A}_i^{\tilde{\sigma}(\sigma-1)} \bar{A}_j^{\tilde{\sigma}\sigma} B_i(\lambda)^{\sigmatil \phi_1} B_j(\lambda)^{\sigmatil \phi_2} e^{ -\frac{\delta}{\beta} \tilde{\sigma}\gamma_2 \lambda_j },
	\end{align}
$i = 1, \dots, n^*, $ together with the aggregate population constraint \eqref{eq:aggpop}.
\end{proposition}
\begin{proof}
    See Appendix \ref{app:proofslambda_eq}.
\end{proof}
The system of equations \eqref{eq:lambda_eq} characterizes the vector of Voronoi weights at a $\Y^*$-centric equilibrium  as a function of the geography and the welfare scalar, incorporating the endogenous shape of the commuting areas as an equilibrium force. Because the weights are defined up to an additive constant, we can redefine them in order to absorb the scalar $V^{(\sigma-1)\frac{\alpha}{\beta}}$ in \eqref{eq:lambda_eq}. If we find a normalized solution, then we can recover the spatial tessellation, and use the aggregate population constraint \eqref{eq:aggpop} to pin down the level of welfare in the economy. The other equilibrium conditions determine the density of residents $\ell$, the wage vector $\{w_i\}_{y_i \in \Y^*}$, and the other components of a $\Y^*$-centric equilibrium.\footnote{Therefore, total population in the economy only affects the level of welfare, as opposed to new economic geography models such as \cite{fujita1995} and \cite{fujita1997}, where changes in the scale of the economy impact on the shape of the equilibrium urban system. In these papers, the source of congestion is an agricultural sector with a Leontief technology that uses labor and land as inputs. In Appendix \ref{app:twosector}, we incorporate an agricultural sector with a Cobb-Douglas technology and we show that the equilibrium system of equations has the same structure as \eqref{eq:lambda_eq}. We conclude that the source of the difference is the specification of the agricultural technology. 

} 

Suppose that we have found a $\Y^*$-centric equilibrium. The next step is to determine if it can be justified as a spatial equilibrium when the full set of business districts is considered. The main challenge is that prices and wages are in general not well-defined when the aggregate labor supply is zero.  Therefore, we conduct the following thought experiment: (i) we consider moving a small group of $L_p > 0$ commuters from their current business district to a vacant one $y_p \in \Y \setminus \Y^*$, (ii) we assess the value of this deviation using the prices and the density of residents prevailing at the current equilibrium; (iii) we study the limit of the market-clearing wage at $y_p$ as $L_p$ approaches zero.  Along the limit, we maintain that the price equal marginal cost condition \eqref{eq:p} holds at $y_p$, while keeping all other prices fixed at their equilibrium values (see Appendix \ref{app:proofslambdap} for more details).

This approach allows us to compute the potential wage associated with an infinitesimal deviation to $y_p $. In line with \cite{fujita2002}, we interpret the potential wage as the highest wage that a zero-profit firm would be prepared to pay at a vacant business district in order to move its production there. Because the price index at vacant business districts is well-defined from equation \eqref{eq:P}, we can also compute the potential Voronoi weight associated with a  business district $y_p$ in $\Y \setminus \Y^*$, 
\begin{equation*}
   \lambda_p = \lim_{L_p \to 0^+} \frac{1}{\delta}  \log \frac{  w_p}{P_p}.
\end{equation*}

We summarize the results in the next proposition. 
\begin{proposition}\label{prop:lambdap}
Suppose that Assumptions \ref{ass:Ab}-\ref{ass:d} hold. For a given $\Y^* \subset \Y$, let $\{\lambda_i\}_{y_i \in \Y^*}$ and $V$ be, respectively, the vector of weights and the level of welfare at a $\Y^*$-centric equilibrium, and let $\lambda_p$ denote the potential weight at a vacant business district $y_p \in \Y \setminus \Y^*$. Then,
\begin{enumerate}[a)]
\item if $\alpha < 1/(\sigma - 1)$, $\lambda_p = +\infty$;
\item if $\alpha > 1/(\sigma - 1)$, $\lambda_p = -\infty$;
\item if $\alpha = 1/(\sigma - 1)$, $\lambda_p$ solves
\begin{equation}\label{eq:lambdap}
       e^{\delta \sigmatil \sigma \lambda_p} =  V^{\frac{1}{\beta}} \sum_{y_j \in \Y^*} T_{pj}^{1-\sigma} \bar{A}_p^{\sigmatil(\sigma-1)}  \bar{A}_j^{\tilde{\sigma}\sigma}  B_j(\lambda)^{-\frac{1}{\beta}} e^{ - \frac{\delta}{\beta} \sigmatil \gamma_2
 \lambda_j }.
\end{equation}
\end{enumerate}
\end{proposition}
\begin{proof}
    See Appendix \ref{app:proofslambdap}.
\end{proof}
This proposition characterizes the incentive to deviate from a $\Y^*$-centric equilibrium as a function of the strength of productivity spillovers, $\alpha$, and the elasticity of substitution across differentiated varieties, $\sigma$.\footnote{In Remark \ref{remark:lamdap}, we also show that, with $\alpha \geq -1$, the aggregate expenditure in $y_p$ always vanishes as $L_p$ approaches zero. }  It anticipates that, unless $\alpha = 1/(\sigma - 1)$, any $\Y^*$-centric equilibrium will either disintegrate ($\alpha < 1/(\sigma - 1)$) or sustain itself as a spatial equilibrium ($\alpha > 1/(\sigma - 1)$) when agents are given the option to relocate to vacant business districts. Because the potential weight is not well-defined in general, in the remainder of this section we first provide sufficient conditions for the existence and uniqueness of a $\Y^*$-centric equilibrium for an \textit{a priori} given urban system $\Y^*$, and then we verify \textit{a posteriori} if such equilibrium is sustainable in face of deviations to vacant business districts.

\subsection{\texorpdfstring{$\Y^*$}{Y*}-centric equilibrium}

Because for $n^* = 1$ a monocentric equilibrium exists and is (trivially) unique, in the next theorem we focus on the case $n^* \geq 2$. 
\begin{theorem}\label{theo:theo1}
Suppose that Assumptions \ref{ass:Ab}-\ref{ass:d} hold and assume that the parameters are such that $\gamma_1 \neq 0$. Then, for any $\Y^* \subseteq \Y$ such that $n^* \geq 2$,
  \begin{enumerate}[a)]
\item there exists $\delta^* > 0$ such that, for $\delta > \delta^*$, a $\Y^*$-centric equilibrium exists. \label{cond:1a}
\item if $| \frac{\gamma_2}{\gamma_1} | < 1$, there exists $\delta^{**} > \delta^{*}$ such that, for $\delta > \delta^{**}$, the $\Y^*$-centric equilibrium is unique. \label{cond:1b}
\end{enumerate}
\end{theorem}
\begin{proof}
    See Appendix \ref{app:proofstheo1}
\end{proof}
This theorem is saying, first, that a $\Y^*$-centric equilibrium may not exist unless $\delta$ is sufficiently high. Intuitively, the condition in part \ref{cond:1a} ensures that real wage disparities never become so large, as compared to commuting costs, as to deter commuters from traveling to one or more business districts in $\Y^*$ (see Lemma \ref{lemma:gmap} in Appendix \ref{app:proofstheo1}). Secondly, a large value of $\delta$ also ensures that, provided that $|\gamma_2 / \gamma_1| < 1$, the following sufficient condition for uniqueness is satisfied:
\begin{equation}\label{eq:suffcondunique} 
    \left\vert \frac{\gamma_2}{\gamma_1} \right\vert +  \sigmatil \bigg(  2 (n^* - 1) \vert \phi_1 \vert  + (2 n^* - 1) \vert \phi_2 \vert  \bigg) \eta < 1,
\end{equation}
where $\eta$ denotes an upper bound for the semielasticity of aggregate residential amenities in a business district $y_i$ with respect to a Voronoi weight $\lambda_k$, with $i \neq k$. This condition is similar to the one obtained in regional models, except for a new term capturing the contribution of the endogenous commuting tessellation. Thus, while $|\gamma_2 / \gamma_1|$ governs the balance between productivity and residential spillovers, $\delta$ ensures that small changes in relative real wages do not result in large swings in the location of the boundaries between commuting areas, so that $\eta$ is small (see Lemma \ref{lemma:etabound} in Appendix \ref{app:proofstheo1}).

The sufficient condition \eqref{eq:suffcondunique} has three main limitations. First, it is not straightforward to verify, because $\eta$ depends on the distance function $d$. Secondly, since tighter estimates are generally not feasible, we must use $n^* -1$, the (potentially large) number of other active sites, as an upper bound on the number of adjacent commuting areas.\footnote{Euler's formula for connected planar graphs implies that the \textit{average} number of neighbors cannot exceed six, but it is not difficult to come up with peculiar geographies such that that the \textit{maximum} number of neighbors exceeds six for certain weight vectors.} Third, while Theorem \ref{theo:theo1} holds in general inside the \textit{open} set $\Lambda$ such that all sites in $\Y^*$ are active, the values of $\delta^*$ and $\delta^{**}$ depend on the choice of an arbitrary \textit{closed} set $\Lambda^c \subset \Lambda$ (see the proof of Proposition \ref{prop:lambda_eq} in Appendix \ref{app:proofslambda_eq} and the proof of Theorem \ref{theo:theo1} in Appendix \ref{app:proofstheo1}).

\subsection{From \texorpdfstring{$\Y^*$}{Y*}-centric equilibria to spatial equilibria}

Which $\Y^*$-centric equilibria are able to to sustain a spatial equilibrium? The next result answers this question.  
\begin{corollary}\label{cor:cor1}
Suppose that the Assumptions \ref{ass:Ab}-\ref{ass:d} hold, and let $\lambda = \{\lambda_i\}_{y_i \in \Y^*}$ be the vector of weights at a $\Y^*$-centric equilibrium. Then,
  \begin{enumerate}[a)]
\item if $\alpha > 1/(\sigma - 1)$, a $\Y^*$-centric equilibrium is always a spatial equilibrium; 
\item if $\alpha < 1/(\sigma - 1)$, no $\Y^*$-centric equilibrium is a spatial equilibrium, unless $\Y^* = \Y$; 
\item if $\alpha = 1 / (\sigma - 1)$, a $\Y^*$-centric equilibrium is a spatial equilibrium if and only if
\begin{equation*}\label{eq:stability}
\sigmatil(\sigma-1) \log \frac{\bar{A}_p}{\bar{A}_i} 
+  \log \frac{\sum_{j = 1}^{n^*} T_{pj}^{1-\sigma} \bar{A}_j^{\tilde{\sigma}\sigma}  B_j(\lambda)^{\sigmatil \phi_2} e^{ -\frac{\delta}{\beta} \tilde{\sigma}\gamma_2 \lambda_j }
	}{\sum_{j = 1}^{n^*} T_{ij}^{1-\sigma} \bar{A}_j^{\tilde{\sigma}\sigma}  B_j(\lambda)^{\sigmatil \phi_2} e^{ -\frac{\delta}{\beta} \tilde{\sigma}\gamma_2 \lambda_j }} 	+ \sigmatil \sigma \delta  d_i(y_p) <0,
 \end{equation*} 
 for all $y_p \in \Y \setminus \Y^*$, with $y_i \in \Y^*$ such that $y_p \in \Omega_i(\lambda)$.
\end{enumerate}
\end{corollary}
This result is a direct consequence of Proposition \ref{prop:lambdap}, therefore we omit the proof. When productivity spillovers are strong (i.e., $\alpha > 1/(\sigma-1)$), the incentive to relocate to a vacant business district is infinitely low. In this scenario, increasing returns generate a lock-in effect at the set of active sites, and any $\Y^*$-centric equilibrium sustains itself as a spatial equilibrium. Conversely, when productivity spillovers are weak (or even negative, i.e., $\alpha < 1/(\sigma -1))$, the incentive to deviate is infinitely high, so that economic activity tends to spread across all available sites. Only in the intermediate case, $\alpha = 1/(\sigma - 1)$, is the potential weight well-defined, and the value of deviating to a vacant business district $y_p \in \Omega_i$ increases with its labor productivity $\bar A_p$, the cost of commuting from $y_p$ to $y_i$, and its trade market access within the extant urban system $\Y^*$. 

While this intermediate case may be of some interest, when $\alpha = 1/(\sigma - 1)$ we are able to prove a stronger result. In fact, in this special case we can represent the spatial equilibria as the solutions to a single system of equations,
\begin{equation*}
       e^{\delta \sigmatil \sigma \lambda_i} =  V^{\frac{1}{\beta}} \sum_{j = 1}^n T_{ij}^{1-\sigma} \bar{A}_i^{\sigmatil(\sigma-1)}  \bar{A}_j^{\tilde{\sigma}\sigma}  B_j(\lambda)^{-\frac{1}{\beta}} e^{ - \frac{\delta}{\beta} \sigmatil \gamma_2
 \lambda_j }, \quad i = 1, \dots, n .
\end{equation*}
Since this system does not depend on $\Y^*$, by studying its solutions we are able to provide a global characterization of the equilibrium properties. 
\begin{theorem}\label{theo:theo2}
Suppose that Assumptions \ref{ass:Ab}-\ref{ass:d} hold, and let $\alpha = 1/(\sigma - 1)$. Then:
  \begin{enumerate}[a)]
\item a spatial equilibrium exists. \label{cond:2a}
\item if $| \frac{\gamma_2}{\gamma_1} | < 1$, there exists $\delta^{***}$ such that, for $\delta > \delta^{***}$, the spatial equilibrium is unique. \label{cond:2b}
\end{enumerate}
\end{theorem}
\begin{proof}
    See Appendix \ref{app:proofstheo2}
\end{proof}
By focusing on a variable that is always positive and well-defined, the theorem abstracts from the fact that some commuting areas may vanish and some empty ones may appear at different values of the vector of weights.\footnote{This result is significant because for $\alpha = 1 / (\sigma - 1)$ the model becomes isomorphic to a monopolistic competition model with fixed costs of production and free entry \citep{allen2014}, the workhorse model in the new economic geography literature.}

Corollary \ref{cor:cor1} indicates that when $\alpha > 1 /(\sigma - 1)$ multiple equilibria exist in the spatial economy. As an illustration, in the next lemma we show that if one urban system meets the sufficient condition for existence, then many other equilibria are generally possible.
\begin{lemma}\label{lemma:multipleY}
Suppose that Assumptions \ref{ass:Ab}-\ref{ass:T2} hold, and consider the sets $\Y^* \subset \Y$ and $\Y^{**} = (\Y^* \setminus \{y_c\} ) \bigcup \{y_p\}$, with  $y_c \in \Y^* $ and $y_p \in \Y \setminus \Y^*$. Further, suppose that $\Y^*$ meets the sufficient condition for the existence of a $\Y^*$-centric equilibrium in Theorem \ref{theo:theo1}, part \ref{cond:1a}. If $ |\bar A_c / \bar A_p |$ and $d_c(y_p) = d_p(y_c)$ are sufficiently small, then there exists a $\Y^{**}$-centric equilibrium.
\end{lemma}
\begin{proof}
    See Appendix \ref{app:proofsmultipleY}
\end{proof}
The lemma is saying that, given the location of the other $n^*-1$ business districts, there is a range of potential sites where the $n^*$-th business districts may sustain an equilibrium -- and thus, for $\alpha > 1/(\sigma - 1)$, a spatial equilibrium.\footnote{\cite{krugman1993} derives an analogous result in a linear economy and for $n^*$ = 1. } More generally, $\Y^*$ and $\Y^{**}$ may differ in the location of multiple sites or the number of sites. 

To summarize, when does a spatial equilibrium exist? The results in this section show that at least one spatial equilibrium must exist when $\alpha \geq 1/(\sigma - 1)$. In this scenario, economic activity may cluster in a limited number of sites. Conversely, when $\alpha < 1/(\sigma - 1)$, the only urban system consistent with a spatial equilibrium is $\Y$. If $\Y$ contains a large number of densely distributed sites, then a spatial equilibrium may fail to exist unless the value of $\delta$ is extremely large. And when is the spatial equilibrium unique? When $\alpha > 1/(\sigma - 1)$, the answer is never: due to indeterminacy in the location of urban centers, the model admits a myriad of spatial equilibria. Conversely, when $\alpha \leq 1/(\sigma - 1)$,  uniqueness is ensured as long as $|\gamma_2 / \gamma_1| < 1$ and $\delta$ is sufficiently large. 

\subsection{Discussion}

\subsubsection{Comparison with the empirical literature}

The empirical literature has documented several facts that have implications for the existence of multiple equilibria in the spatial economy: (i) cities preserve their location even when their natural advantages become obsolete \citep{bleakely2012}; (ii) following the  ``reset'' of an urban system, urban location may change \citep{michaels2018}; (iii) cities converge back to their long-run relative sizes following large negative shocks \citep{davis2002};\footnote{\cite{brakman2004} reach the same conclusion for German cities after the World War II bombings. However, in follow-up studies \cite{bosker2007} and \cite{bosker2008} find evidence of multiple equilibria in the German city-size distribution. } . While facts (i) and (ii) support a model where increasing returns dominate geographic advantages, leading to multiple equilibria and path-dependence, fact (iii) is usually interpreted as evidence in support of a model where the equilibrium is unique, geographic advantages drive the distribution of economic activity, and increasing returns to scale play a secondary role. 

Our theoretical results show that our framework can provide a unified explanation for all these facts. The key insight is that the model allows for \textit{multiple} equilibria in the location of active business districts but a \textit{unique} distribution of labor across them. A necessary condition for this outcome is $\alpha > 1/(\sigma -1)$. Moreover, if $|\gamma_2 / \gamma_1 | < 1$ and $\delta$ is large, then this is the only possible scenario for any geography -- see Theorem \ref{theo:theo1} and Corollary \ref{cor:cor1}. In Theorem \ref{theo:theo3}, presented in the next subsection, we also show that the same scenario arises when the distance between urban centers is sufficiently large.

For $\alpha > 1/(\sigma - 1)$, historical contingencies that cause the emergence of an urban center will have persistent effects. Once an urban center is established, increasing returns to scale lock in economic activity, preventing the local labor force from abandoning it in favor of a vacant site, even when the original episode becomes irrelevant (Fact (i)).

To ``switch off'' a city, the model requires an event that wipes it out entirely. Even then, Lemma 6.1 suggests that if a new urban center were to replace the old one, it may form not far from its predecessor. Conversely, when a catastrophic event causes the collapse of the entire urban system, then more possibilities arise, and the indeterminacy in the location of cities might be resolved in novel ways (Fact (ii)).

In the absence of events capable of causing discrete changes in the structure of the urban system, the model predicts that even dramatic temporary shocks to city sizes do not alter the spatial equilibrium when $|\gamma_2 / \gamma_1| < 1$ and $\delta$ is large (Fact (iii)). Moreover, at this spatial equilibrium, the city-size distribution is governed by equation \eqref{eq:lambda_eq}, consistent with the empirical evidence on the effect of natural advantages on urban size. 

\begin{figure}
\caption{Multiple spatial equilibria and model parameters}
\centering
\begin{subfigure}{0.45\textwidth}
\centering
\caption{Multiplicity of urban location: $\alpha$ and $\sigma$}
\includegraphics[width=\textwidth]{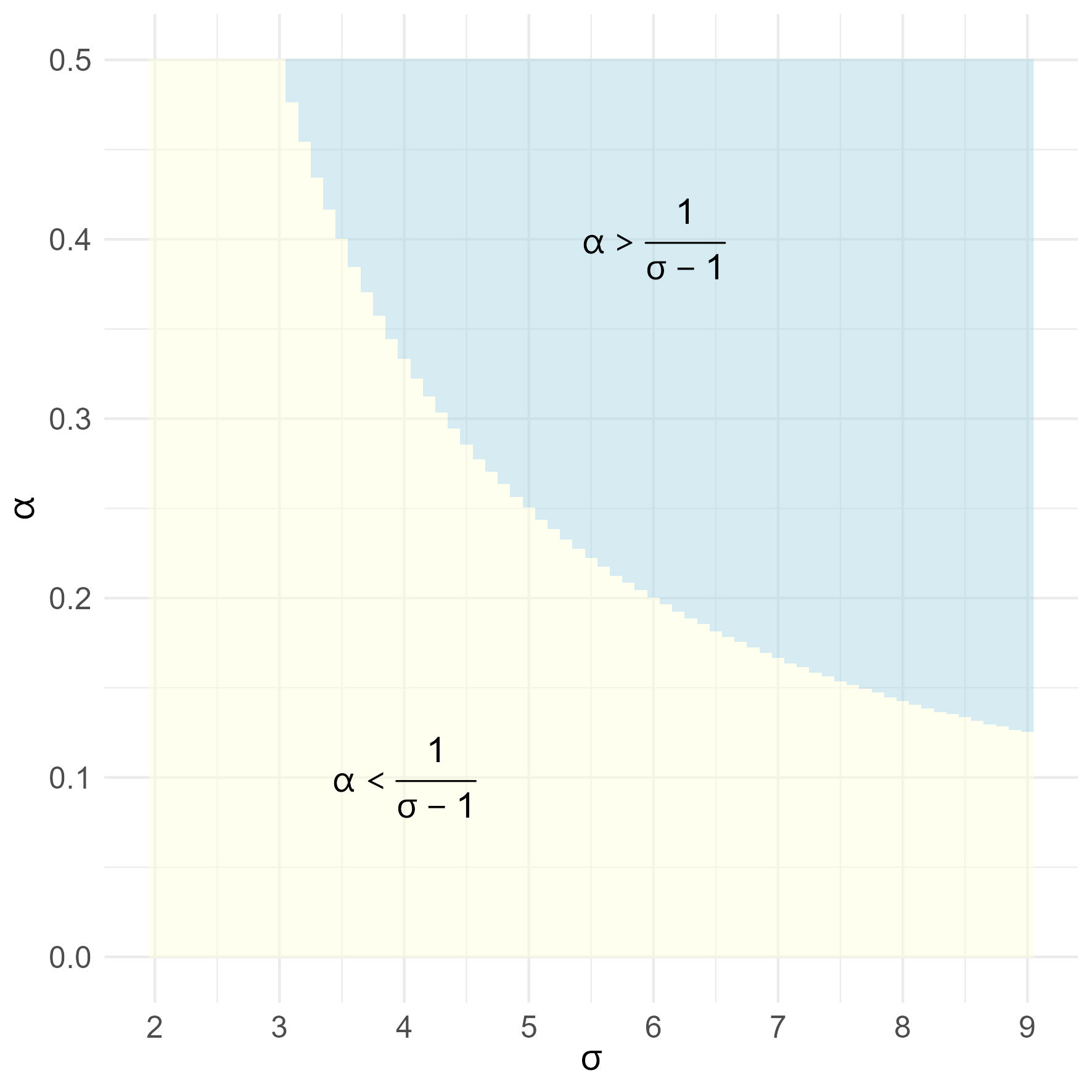}
\label{fig:alphasigma}
\end{subfigure} 
\\
\begin{subfigure}{0.45\textwidth}
\caption{Multiplicity of urban location and uniqueness of the labor distribution: $\sigma = 5$}
\includegraphics[width=\textwidth]{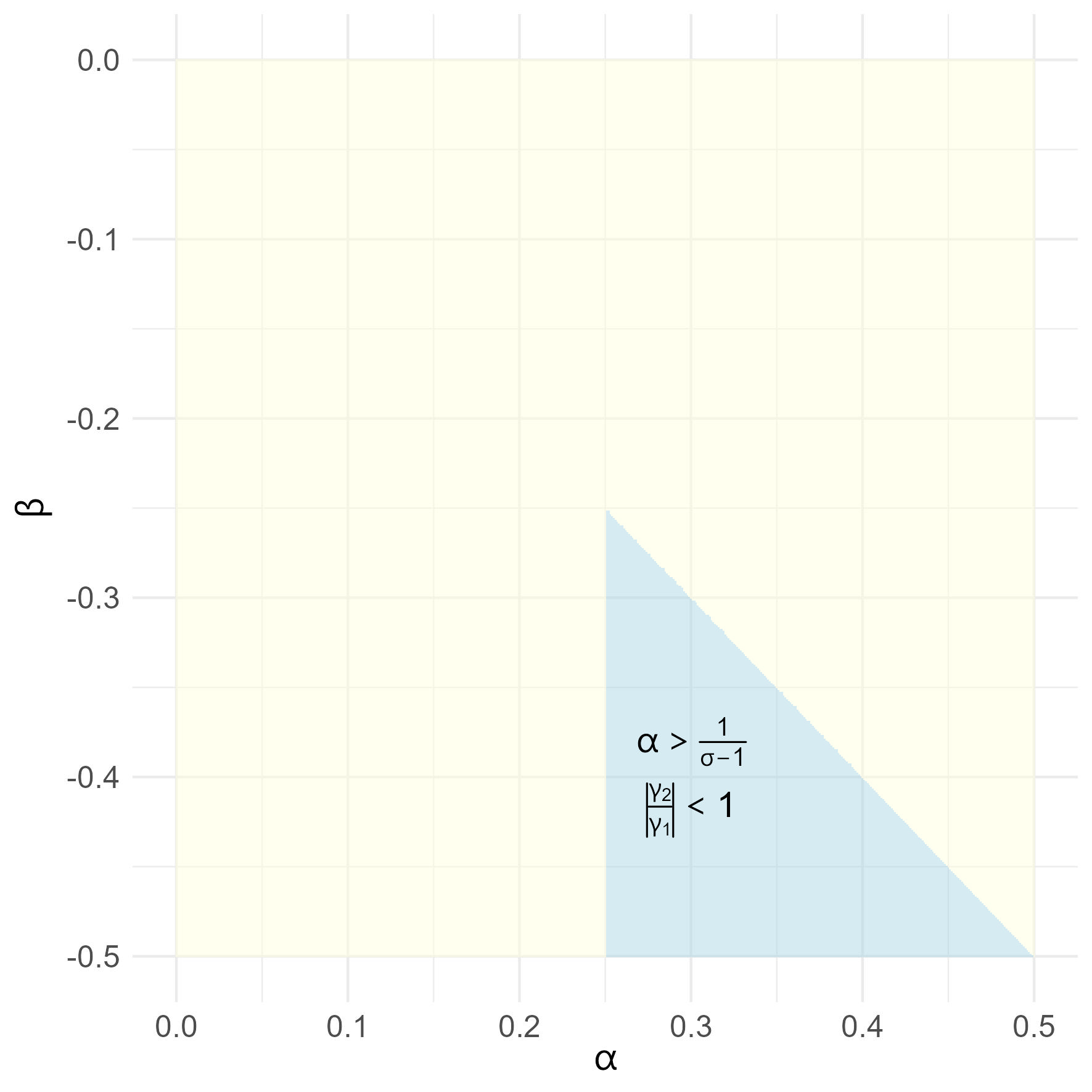}
\label{fig:alphabetasigma5}
\end{subfigure} 
\begin{subfigure}{0.45\textwidth}
\caption{Multiplicity of urban location and uniqueness of the labor distribution: $\sigma = 9$}
\includegraphics[width=\textwidth]{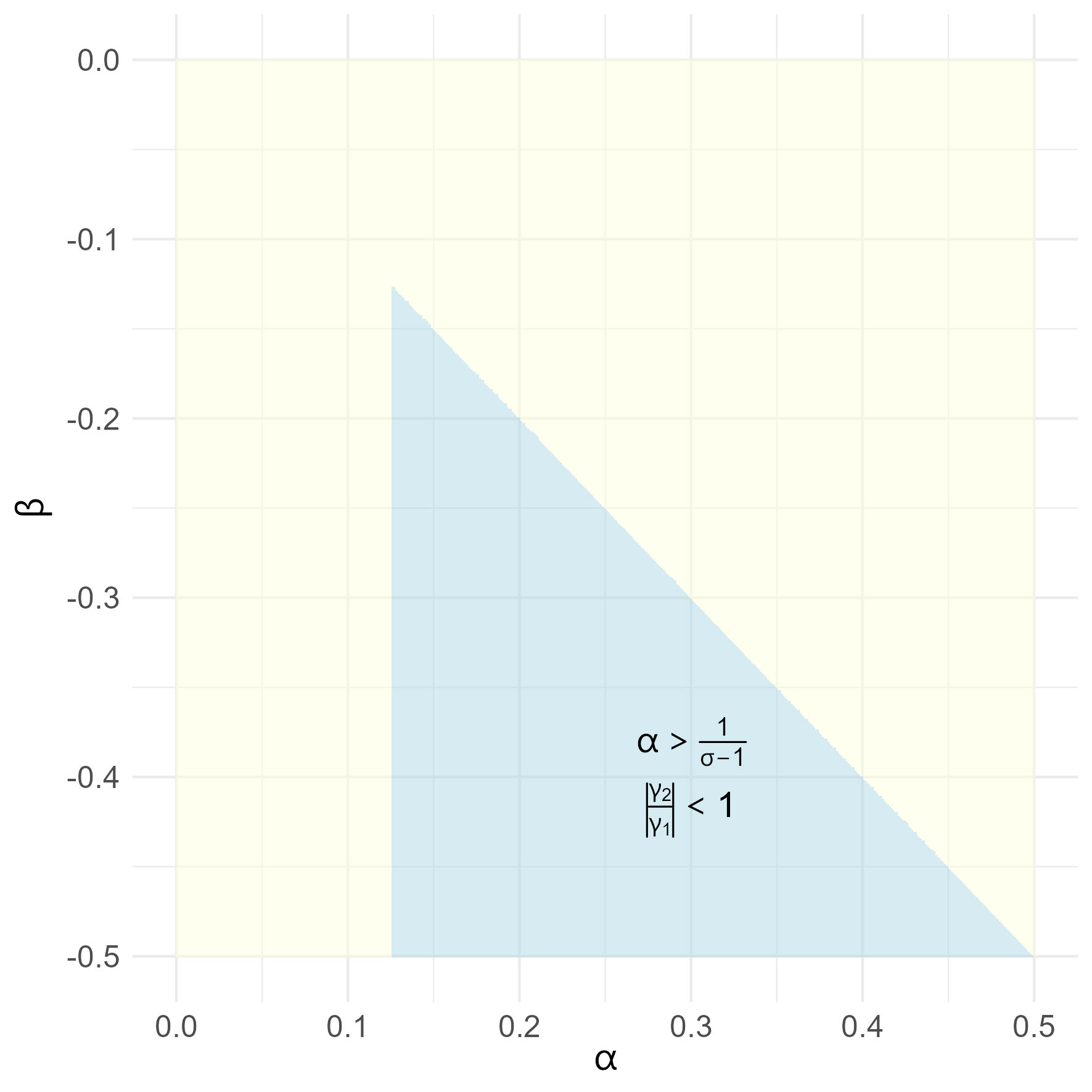}
\label{fig:alphabetasigma9}
\end{subfigure} 
\label{fig:alphabetasigma}
\end{figure}

Are these parameter values plausible? In Figure \ref{fig:alphabetasigma}, we show different combinations of $\alpha$, $\beta$ and $\sigma$ along with the corresponding regions of the parameter space. Figure \ref{fig:alphabetasigma} illustrates that values of $\alpha$ and $\sigma$ such that $\alpha > 1 / (\sigma - 1)$ fall well within the range of the estimates available in the literature. Moreover, Figures \ref{fig:alphabetasigma5} and \ref{fig:alphabetasigma9} show that this condition agrees with $|\gamma_2 / \gamma_1|<1$ in a plausible region of the parameter space. For instance, these conditions are satisfied for $\alpha = 0.2$, $\beta = -0.3$, and $\sigma = 9$. While it is difficult to pin down the value of $\delta$ that satisfies the sufficient condition in Theorem \ref{theo:theo1}, part \ref{cond:1b}, we conclude that multiple equilibria in urban location are not inconsistent with a unique labor distribution across active business districts.

In sum, a scenario where multiple equilibria in the location of cities coexist with a unique distribution of labor across them may reconcile seemingly conflicting facts on the existence of multiple equilibria in the spatial economy and is consistent with the parameter estimates available in the empirical literature.

\subsubsection{The spatial structure of urban systems}

In Theorem \ref{theo:theo1}, we have fixed $\Y^*$ and characterized $\Y^*$-centric equilibria in terms of the model parameters, emphasizing the role of $\delta$. A complementary exercise is to ask how the equilibrium properties depend on the spatial structure of the urban system for a given value of $\delta$. The motivation is twofold. First, when $\alpha > 1/(\sigma - 1)$, all $\Y^*$-centric equilibria are spatial equilibria that we may expect to emerge in the real world. Second, in certain research settings, the researcher may be willing to fix the urban system in advance, so that the relevant equilibrium concept is the $\Y^*$-centric equilibrium. In the next theorem, we provide an answer to this question under the additional assumption that all distances are determined by a common metric.
\begin{assumption}\label{ass:dmetric}
    For all $y_i \in \Y$ and $x \in \X$, $d_i(x) = d(y_i, x)$ where $d$ is a distance function.
\end{assumption}
We focus on one specific summary statistic of urban systems, the minimum distance between its business districts, $    d_{min} = \min_{ \substack{y_i, y_j \in \Y^* \\ y_i \neq y_j}} d(y_i, y_j)$.

\begin{theorem}\label{theo:theo3}
Suppose that Assumptions \ref{ass:Ab}-\ref{ass:dmetric} hold and assume that the parameters are such that $\gamma_1 \neq 0$, and
\begin{equation*}
    -\frac{\delta}{\beta}\sigmatil |\gamma_1| > \tau(\sigma - 1) 
\end{equation*}
Then, for any $\Y^* \subseteq \Y$ such that $n^* \geq 2$,
\begin{enumerate}[a)]
\item there exists $d^* > 0$ such that, for all $\Y^{*} \subseteq \Y$ with $d_{min} > d^*$, a $\Y^*$-centric equilibrium exists. \label{cond:3a}
\item if $| \frac{\gamma_2}{\gamma_1} | < 1$, there exists $d^{**} > d^{*}$ such that, for all $\Y^{*} \subseteq \Y$ with $d_{min} > d^{**}$, the $\Y^*$-centric equilibrium is unique. \label{cond:3b}
\end{enumerate}
\end{theorem}
\begin{proof}
    See Appendix \ref{app:proofstheo3}.
\end{proof}
The basic intuition for this theorem is similar to the one of Theorem \ref{theo:theo1}, except that by considering different urban systems, we are not only altering commuting costs but also potential differences in trade market access across business districts. Under the stated restriction on $\delta$ and $\tau$, required to ensure existence, the commuting cost component prevails. Therefore, as business districts are spaced further apart, they become more insulated from competition for commuters, and  commuting areas become less responsive to economic incentives. Note that, because $\X$ is bounded, the requirement that the distances between business districts exceed a certain cutoff may fail for all $\Y^* \subseteq \Y$. 

One way to interpret the choice of $\Y^*$ is from the viewpoint of a researcher deciding on the appropriate discretization of the spatial economy, without imposing predetermined boundaries between spatial units. As an example, depending on the research question, the set $\Y^*$ might include county seats, central business districts within metropolitan areas, or business districts within individual cities. Then, Theorem \ref{theo:theo3} is saying that equilibrium existence and uniqueness are guaranteed as long as $|\gamma_2 / \gamma_1|<1$ and the researcher is willing to adopt a sufficiently coarse view of the spatial economy. As the distance between the selected sites increases, the model approaches a regional model, except that the boundaries between spatial units do not coincide with political or administrative boundaries, but will instead correspond to a standard Voronoi tessellation. 

\subsubsection{The role of geography}

Because variations in geographic characteristics translate into differences in real wages, the sufficient conditions for the existence of a $\Y^*$-centric equilibrium in Theorems \ref{theo:theo1} and \ref{theo:theo3} depend on the underlying geography. This observation naturally raises the question: which geographies, and which spatial arrangements of urban centers within them, are more conducive to the existence of an equilibrium urban system?

To address this question, we revisit the approach used in the proofs of Theorems \ref{theo:theo1} and \ref{theo:theo3}. In both cases, the proof relies on the fact that when either $\delta$ or $d_{\text{min}}$ is large, the specific location of business districts within the geographic space $\X$ becomes negligible. To capture this aspect of the problem, in the next lemma we use the standard Voronoi tessellation (i.e., $\lambda = 0$) as a benchmark to isolate the exogenous component of aggregate residential amenities. As a preliminary step, define
\begin{align*}
    r = \min_{y_i \in \Y^*} \max_{y_j \in \Y^*} d_i(y_j)
\end{align*}
\begin{lemma}\label{lemma:gmap2}
Suppose that Assumptions \ref{ass:Ab}-\ref{ass:T2} hold. Then, for any $\Y^* \subseteq \Y$ such that $n^* \geq 2$, if
\begin{equation}\label{eq:gmap2}
    \sigmatil(\sigma - 1) \left| \log \frac{\bar A_i}{\bar A_{j}} \right| + \sigmatil |\phi_1| \left|\log  \frac{B_i(0)}{B_{j}(0)} \right| - 2 \beta \eta r  \leq \left( - \frac{\delta}{\beta} \sigmatil |\gamma_1| -  \tau (\sigma - 1) \right) d_i(y_{j})
\end{equation}
for all $y_i,y_j \in \Y^*$ , $y_i\neq y_j$, a $\Y^*$-centric equilibrium exists. 
\end{lemma}
\begin{proof}
    See Appendix \ref{app:proofsgmap2}
\end{proof}
While the right-hand side of condition \eqref{eq:gmap2} captures the impact of bilateral distances across business districts on commuting and trade market access, as discussed in Theorem \ref{theo:theo3}, the left-hand side summarizes the role of other exogenous features of the geography and the selected urban system. First, the terms $|\bar A_i / \bar A_j|$ reflect labor productivity differences across sites. An urban system is more likely to satisfy the sufficient condition for existence when these differences are not too pronounced, as a site with very poor labor productivity compared to a nearby one will struggle to attract commuters.

Second, the terms $|B_i(0) / B_j(0)|$ encode the spatial distribution of business districts within the economic space $\X$ and its exogenous measure of residential amenities. In settings where residential amenities are uniform, these terms become smaller when business districts are evenly distributed over $\X$, each of them serving an approximately equal commuting area in the standard tessellation. Conversely, when residential amenities vary across locations, differences in aggregate residential amenities are minimized when areas with richer amenities host a denser concentration of business districts. 

Third, and finally, term $r$ captures two features of urban systems: first, their spatial extension, and, second, the presence of a business district in a central position. If $\Y^*$ spans a large area, it will allow for larger real wage differentials across business districts, making it more difficult to meet the inequalities in \eqref{eq:gmap2}. However, the presence of a business district in a central position lowers $r$ and provides a sharper bound to (normalized) real wage differentials, thus relaxing the sufficient condition.

\section{Concluding remarks}

In this paper, we have revisited the question of multiple equilibria in economic geography through the lens of a quantitative model of a spatial urban system, featuring a continuum of residential locations and a finite set of sites for production and trade. In this setting, both the set of active business districts and the shape of their commuting areas are determined in equilibrium. This spatial structure allows us to bridge  new economic geography models of systems of cities and regional models. On the one hand, when increasing returns to scale are sufficiently strong, the introduction of endogenous commuting areas revives the tendency toward multiple equilibria found in new economic geography models, due to the indeterminacy in urban location. On the other hand, if the urban system is taken as given and the business districts are placed sufficiently far apart, the model converges toward a regional model with discrete locations. Furthermore, for plausible parameter values, the model yields multiple equilibria in urban location and a unique equilibrium in the size of urban centers, conditional on the urban system. This scenario provides a unified explanation for several empirical findings that are often seen as offering conflicting evidence on the existence of multiple equilibria in the spatial economy.

Our analysis opens up several directions for future research. First, dynamic extensions of our framework may deliver insights into which equilibrium urban systems are selected out of the myriad of possible spatial equilibria. Second, it would interesting to extend the model to incorporate multiple sectors of production, subject to different degrees of increasing returns and transportability.

	\newpage
	\bibliographystyle{chicago}

	\newpage

\appendix

\section{Mathematical Tools}

\setcounter{equation}{0}
\renewcommand{\theequation}{A.\arabic{equation}} 

\subsection{Additively weighted Voronoi tessellations}\label{app:voronoi}

Let $\X$ be an open, bounded, connected subset of the Euclidean plane $\R^2$ with Lipschitz boundary $\partial X$, let $\Y \subset \X$ be a finite set of $n$ points, and for each $y_i \in \Y$, let
\[
d_i\colon  \R^2 \to \R_+
\]
be a continuous function that assigns to a point $x \in \R^2$ a nonnegative value $d_i(x)$. We refer to the system $\{d_i(\cdot)\}_{i=1}^n$ as the system of ``distance'' functions.

A spatial tessellation is a partition of $\X$ in a collection of regions without gaps or overlaps. A standard Voronoi tessellation is a spatial tessellation of $\X$ in a collection of regions $\{\Omega_i\}_{i=1}^n$ such that each $x$ is assigned to the nearest $y_i$, i.e.,
\begin{equation*}
x \in \Omega_i \iff    d_i(x) < \min_{i \neq j} d_j(x).
\end{equation*}
Now, assume that each point $y_i \in \Y$ is associated with a weight $\lambda_i \in \R$, and let $\lambda = (\lambda_1, \dots, \lambda_n)$ be the vector of all weights, for an arbitrary ordering $y_1, \dots, y_n \in \Y$. An additively weighted Voronoi tessellation is a spatial tessellation such that each $x$ is assigned to the nearest $y_i$ based on the weighted distance $d_i(x) - \lambda_i$, i.e.,
\begin{equation}\notag
x \in \Omega_i \iff d_i(x) - \lambda_i < \min_{j \neq i} d_j(x) - \lambda_j.
\end{equation}
Any two vectors $\lambda \neq \lambda^{\prime}$ such that the weights differ only by an additive constant generate the same additively weighted Voronoi tessellation. Any vector $\lambda$ such that all weights are equal to the same constant generates a standard Voronoi tessellation. 

 To construct an additively weighted Voronoi diagram more formally, we define, for $y_i\neq y_j \in \Y$ and $c \in \R$, the bisector
\begin{equation*}\label{eq: bisector}
B_c(y_i,y_j) = \{x \in \R^2\colon d_i(x)-d_j(x) = c\},
\end{equation*}
and the dominance region
\begin{equation*}\label{eq: dom}
R_c(y_i,y_j) = \{x \in \R^2\colon d_i(x) - d_j(x) < c \}.
\end{equation*}
Then, we say that $B_{\lambda_i-\lambda_j}(y_i,y_j)$ is the additively weighted bisector of $y_i$ and $y_j$ and that $R_c(y_i,y_j)$ is the additively weighted dominance region of $y_i$ over $y_j$. Note that, in general, the dominance regions span the entire plane. We want to use them to create a tessellation of $\X$. We say that
\begin{equation}\label{eq: voronoi}
\Omega_i (\lambda) = \X \cap \bigcap_{j\neq i}R_{\lambda_i-\lambda_j}(y_i,y_j)
\end{equation}
is the additively weighted Voronoi region of $y_i$ (with respect to $\Y$) in $\X$. Finally, the additively weighted Voronoi tessellation of $\Y$ in $\X$ is defined as
\[
\bigcup_{i=1}^n \bar \Omega_i(\lambda).
\]

Following \cite{geiss2013}, we impose a technical condition on the system of distance functions $\{d_i\}_{y_i\in \Y}$ to ensure that the Voronoi tessellations are well-behaved. 

\begin{definition}\label{def:distance}
A system of continuous distance functions $d_i(\cdot)$, for $y_i \in \Y$, is called \textit{admissible} if for all $y_i\neq y_j \in \Y$, and for each bounded open set $C \subset \R^2$, there are two constants $m_{ij}$ and $M_{ij}$, with $m_{ij} < M_{ij}$, such that $c \mapsto |C \cap R_c(y_i,y_j)|$ is continuously increasing from $0$ to $|C|$, the Lebesgue measure of $C$, as $c$ grows from $m_{ij}$ to $M_{ij}$. Moreover, $C \cap R_c(y_i,y_j) = \emptyset$ if $c \leq m_{ij}$, and $C \subset R_c(y_i,y_j)$ if $c \geq M_{ij}$.
\end{definition}

Under this assumption, the bisectors have measure $0$.

\begin{lemma}[\cite{geiss2013}] \label{lem: bisectors} 
For a system of admissible distance functions $d_i(\cdot)$, where $y_i \in \Y$, for any two points $y_i\neq y_j \in \Y$ and any $c \in \R$, and for any bounded open set $C \subseteq \R^2$, we have $|C\cap B_c(y_i,y_j)|=0$. 
\end{lemma}
Above, we have constructed Voronoi regions as the intersections of dominance regions. This approach clarifies that the boundary of a Voronoi region $\Omega_i$ consists of the union of specific segments lying along the bisectors between $\Omega_i$ and its neighboring regions. In the next subsection, we will use this fact to compute the partial derivative of an integral over $\Omega_i$ with respect to a Voronoi weight.

\subsection{Shape derivative}\label{app:shapeder}
Let $F\colon \R^{n} \to \R$ be defined as
\begin{equation*}
    F(\lambda) = \int_{\Omega_i(\lambda)} f(x, \lambda) \d x,
\end{equation*}
for some real-valued differentiable function $f$. We are interested in the partial derivative of $F$ with respect to a Voronoi weight $\lambda_k$, $k = 1, \dots, n$. In \cite{lanzara2023}, we derive a formula for this derivative by applying classical results in the theory of shape optimization to the specific context of Voronoi tessellations (see, in particular, Theorem 5.2.2 in \cite{henrot2006}). We have
\begin{align}\label{eq:shapeder}
    \frac{\partial}{\partial \lambda_k} \int_{\Omega_i(\lambda)} f(x, \lambda) \d x  &= 
    \begin{dcases} \int_{\Omega_i(\lambda)} \frac{\partial}{\partial \lambda_k} f(x, \lambda) \d x + S_{ik}, \quad k \neq i , \\
    \int_{\Omega_i(\lambda)} \frac{\partial}{\partial \lambda_i} f(x, \lambda) \d x + \sum_{\ell \in \NN_i} S_{i \ell},  \quad k = i ,
    \end{dcases}\\ \notag
    \text{ with } S_{ik} &\equiv \int_{\Gamma_{ik}}f(x)  \frac{\partial \omega_{ik}}{\partial \lambda_k} (x,\lambda) \cdot \nu_i(x) d\sigma(x),
\end{align}

and where
\begin{itemize}
\item  $\NN_i$ is the set of neighbors of $y_i$; that is, \[\NN_i = \{k = 1,\dots,n, k \neq i : \partial \Omega_k \cap \partial \Omega_i \neq \emptyset\},\]
\item $\Gamma_{ik}$ is the border segment between $\Omega_i$ and $\Omega_k$; that is, \[ \Gamma_{ik}\subset \partial \Omega_i \text{ with } \cup_{k \in \NN_i}\Gamma_{ik} = \partial \Omega_i \text{ and }\partial \Omega_k \cap \partial \Omega_i = \Gamma_{ik},\]
\item the map $(x,\lambda) \mapsto \omega_{ki}(x,\lambda)$ is a local parametrization of the curve $\Gamma_{ik}$,
\item $\nu_i(x)$ is the unit normal vector pointing outside of $\Omega_i$ at $x \in \partial \Omega_i$,
\item $d\sigma(x)$ is the surface measure on $\partial \Omega_i$.
\end{itemize}
Equation \eqref{eq:shapeder} is a generalization of the Leibniz rule. It says that the derivative of $F$ with respect to $\lambda_k$ has two components. The first component accounts for the change in the value of $f$ inside $\Omega_i$, keeping $\Omega_i$ fixed. The second component accounts for change in the domain of integration, keeping $f$ fixed. As $\lambda_k$ varies, for $k\neq i$, only the border segment between $\Omega_i$ and $\Omega_k$ is affected. If $\Omega_i$ and $\Omega_k$ are not contiguous, then this component of the derivative is zero.  As $\lambda_i$ varies, all border segments between $\Omega_i$ and its neighbors are affected at the same time.  

\section{Proofs}\label{app:proofs}

\setcounter{equation}{0}
\renewcommand{\theequation}{B.\arabic{equation}} 

\subsection{Proof of Proposition \ref{prop:lambda_eq}} \label{app:proofslambda_eq}

\begin{proof}[Proof of Proposition \ref{prop:lambda_eq}]
For a given commuting pattern $\{\Omega_i\}_{y_i \in \Y^*}$ and $\ell$, such that $|\Omega_i| > 0$ for all $y_i \in \Y^*$, where $|\cdot|$ denotes the Lebesgue measure, we can reduce equations \eqref{eq:C}-\eqref{eq:lmc} to the following system of equations:
\begin{align}\label{eq:marketeq}
w_i^{\sigma} L_i &= \sum_{j = 1}^{n^*}  T_{ij}^{1 - \sigma} A_i^{\sigma - 1}   P_j^{\sigma - 1}  w_j L_j, \\ \label{eq:marketeq2}
P_i^{1-\sigma} &= \sum_{j = 1}^{n^*} T_{ji}^{1-\sigma} A_j^{\sigma - 1} w_j^{1-\sigma},
\end{align}
for $i = 1,\dots, n^*$. This system defines a market equilibrium \citep{allen2014, allen2020}. 

Next, agents choose a commuting destination and a residential location to maximize their indirect utility \eqref{eq:V}. First, the choice over $\Y^*$ in the commuting problem \eqref{eq:maxV} implies that we can represent the partition of $\X$ into  commuting areas as an additively weighted Voronoi tessellation, where the Voronoi weights are given in equation \eqref{eq:lambda} and the Voronoi regions are given in equation \eqref{eq:Omega}. Under Assumption \ref{ass:trineq} (triangle inequality), we can define the unbounded open set $\Lambda \subset \R^{n^*}$, 
\begin{equation*}
 \Lambda = \{ \lambda \in \R^{n^*}: 
- d_j(y_i) < \lambda_i - \lambda_j < d_i(y_j),  \text{ all } i,j = 1,\dots, n^*,  i \neq j \},
\end{equation*} 
such that $\lambda \in \Lambda$ if and only if all business districts in $\Y^{*}$ attract a positive measure of commuters in the corresponding additively weighted Voronoi tessellation. 

Second, with $\beta < 0$, the choice over $\X$ implies that all residential locations host a positive measure of agents and yield the same level of welfare at equilibrium, that is, $V(x, y_i) = V(x^{\prime}, y_j) = V$ for all $x, x^{\prime} \in \mathcal{X}$ and $y_i, y_j \in \Y^{*}$, such that $x \in \Omega_i$ and $x^{\prime} \in \Omega_j$, where the scalar $V>0$ is the common level of welfare in the economy. By equations \eqref{eq:b}, \eqref{eq:lmc}, and \eqref{eq:V}, welfare equalizes inside a nonempty commuting area $\Omega_i$ if and only if the density of residents is equal to
\begin{align}\label{eq:l}
\ell_i(x) &= \frac{\left( \bar{b}(x)/D(x, y_i)\right)^{-\frac{1}{\beta}}}{\int_{\Omega_i}\left( \bar{b}(x)/D(x, y_i)\right)^{-\frac{1}{\beta}}\d x } L_i(\lambda) \quad x \in \Omega_i , y_i \in \Y^*,
\end{align}
with $\ell(x) = \sum_{i = 1}^{n^*} \chi_{\Omega_i}(x) \ell_i(x)$. Using this expression into the indirect utility \eqref{eq:V}, we can write the equalized level of welfare within $\Omega_i$ as in equation \eqref{eq:Vi}, and the welfare equalization condition across business districts as equation \eqref{eq:weq}.

Fix a vector of weights $\lambda \in \Lambda$. The market equilibrium conditions \eqref{eq:marketeq} and \eqref{eq:marketeq2}, together with the welfare equalization condition, \eqref{eq:weq} deliver the same set of equilibrium equations as in \cite{allen2014}. Under Assumptions \ref{ass:Ab}, they show that the following relationship holds for all $i = 1, \dots, n$: 
\begin{equation}\label{eq:wscale}
    w_i^{\sigma} \bar{A_i}^{1 - \sigma} L_i(\lambda)^{1 - \alpha(\sigma-1)} = c w_i^{1-\sigma} B_i(\lambda)^{1- \sigma} L_i(\lambda)^{ \beta(1-\sigma)},
\end{equation}
where $c$ is a positive constant. Furthermore, they show that we can combine the equilibrium conditions into a single system of equations,
\begin{align}
	\label{eq:L} L_i(\lambda)^{\tilde{\sigma}\gamma_1} &= V^{1-\sigma} \sum_{j = 1}^{n^*} T_{ij}^{1-\sigma} \bar{A}_i^{\tilde{\sigma}(\sigma-1)} \bar{A}_j^{\tilde{\sigma}\sigma} B_i(\lambda)^{\tilde{\sigma}\sigma} B_j(\lambda)^{\tilde{\sigma}(\sigma-1)} L_j(\lambda)^{\tilde{\sigma}\gamma_2}, \\ \notag
i = 1,\dots,n^*,  \text{ with }   \gamma_1 &= 1 - (\sigma - 1) \alpha - \sigma \beta, \quad 
	\gamma_2 = 1 + \sigma \alpha + (\sigma -1 ) \beta,  \quad \notag
	\sigmatil = \frac{\sigma-1}{2\sigma - 1} .
	\end{align}
 This system characterizes the equilibrium distribution of labor supply across commuting areas when the spatial tessellation is given. As a final step, we need to incorporate the choice of a commuting destination. Using \eqref{eq:lambda} and \eqref{eq:weq} into \eqref{eq:L}, we obtain 	
\begin{align*}
	e^{-\frac{\delta}{\beta} \tilde{\sigma}\gamma_1 \lambda_i } &= V^{(\sigma-1){\frac{\alpha}{\beta}}} \sum_{j = 1}^{n^*} T_{ij}^{1-\sigma} \bar{A}_i^{\tilde{\sigma}(\sigma-1)} \bar{A}_j^{\tilde{\sigma}\sigma} B_i(\lambda)^{\sigmatil \phi_1} B_j(\lambda)^{\sigmatil \phi_2} e^{ -\frac{\delta}{\beta} \tilde{\sigma}\gamma_2 \lambda_j }, \\ \notag
 \text{for } i &= 1,\dots, n^*, \text{ with }
	\phi_1 = \frac{(1-(\sigma-1)\alpha )}{\beta},
	\phi_2 =  - \frac{(1 + \sigma\alpha ) }{\beta}. \qedhere
	\end{align*}
\end{proof}

\subsection{Proof of Proposition \ref{prop:lambdap}}\label{app:proofslambdap}

Consider moving $L_p$ commuters to $y_p \in \Y \setminus \Y^*$, with $y_p \in \Omega_i(\lambda)$, while keeping all prices fixed at their equilibrium values. Such arrangement may benefit both firms and commuters whenever it is possible to find a price $p_p$ and a wage $w_p$ such that firms earn positive profits and commuters receive a higher level of welfare, i.e., $w_p < w^*_p = p_p A_p$,  and $V(x, y_p) > V(x, y_i)$, or 
\begin{align*}
    \frac{w_i}{D(x,y_i)P_i} <& \frac{w_p}{D(x, y_p)P_p} < \frac{w^*_p}{D(x, y_p)P_p} \iff \\ 
    \lambda_i - d_i(x) &< \lambda_p - d_p(x) 
\end{align*}
for some $x \in \X$, where $w^*_p$ is the wage rate that makes firms break even, and
\begin{equation*}
    \lambda_p = \frac{1}{\delta} \log \frac{w^*_p}{P_p},
\end{equation*}
in analogy with equation \eqref{eq:lambda}. Under the triangle inequality (Assumption \ref{ass:trineq}), $V(y_p, y_p) \leq V(y_p, y_i)$ implies $V(x, y_p) \leq V(x, y_i)$ for all $x \in \X$.  As a result, a $\Y^*$-centric equilibrium will be sustainable as a spatial equilibrium if and only if 
\begin{equation*}
    \lambda_p < \lambda_i - d_i(y_p), \quad \text{for all } y_p \in \Y \setminus \Y^*. 
\end{equation*}
 
As a final step, then, we define the potential wage $w^*_p$ as the limit of the market-clearing wage as $L_p$ approaches zero. Formally, for $L_p$ such that $0<L_p < \min_{y_i \in \Y^*} L_i $ let $L_j^- = L_j$ for all $j \neq i$ and $L_i^- = L_i - L_p$. We say that
\begin{align*}
    w^*_p &= \lim_{L_p \to 0^+} w_p,
    \text{ such that $w_p > 0$ solves } \\ w_p^\sigma L_p^{1-\alpha(\sigma-1)}   &= \sum_{j = 1}^{n^*} T_{pj}^{1-\sigma} \bar{A}_p^{\sigma-1} P_j^{\sigma-1} w_j L^{-}_j + \bar{A}_p^{\sigma-1}P_p^{\sigma-1} w_p L_p.
\end{align*}

The next lemma shows that $w^*_p$ is well-defined. 
\begin{lemma}
Let the function $f\colon \R_{+} \to \R$ be defined from 
\begin{equation*}
    f(w) = w^\sigma L_p^{1-\alpha(\sigma-1)} - \bar{A}_p^{\sigma-1}P_p^{\sigma-1} w L_p  - \sum_{j = 1}^{n^*} T_{pj}^{1-\sigma} \bar{A}_p^{\sigma-1} P_j^{\sigma-1} w_j L^{-}_j.
\end{equation*}
Then, $f(w) = 0$ has a unique positive solution for $L_p > 0$. 
\end{lemma}
\begin{proof}
We have that 
\begin{equation*}
f(0) = - \sum_{j = 1}^{n^*} T_{pj}^{1-\sigma} \bar{A}_p^{\sigma-1} P_j^{\sigma-1} w_j L^{-}_j < 0, \quad \lim_{w \to +\infty} f(w) = +\infty,
\end{equation*}
and $f^{\prime}(w) = \sigma w^{\sigma - 1}L_p^{1-\alpha(\sigma-1)} - \bar{A}_p^{\sigma-1}P_p^{\sigma-1} L_p $, so that 
\begin{equation*}
    f^{\prime}(w) > 0 \iff w > w_{min} \equiv \left(\frac{1}{\sigma}\right)^{\frac{1}{\sigma-1}} \bar{A}_p P_p L_p^{\alpha} > 0.
\end{equation*}
Because $w_{min}$ is a global minimum, $f(0) < 0 \implies f(w_{min}) < 0$. It follows that there exists a unique $w_p > w_{min}$ such that $f(w_p) = 0$. 
\end{proof}
With these preliminary definitions in hand, we now prove Proposition \ref{prop:lambdap}. 
\begin{proof}[Proof of Proposition \ref{prop:lambdap}]
Consider the case $n^*< n$, such that $\Y \setminus \Y^*$ is nonempty. For $L_p >0$, consider the family of functions $f^{L_p}: \R_{+} \to \R$ with
\begin{equation*}
    f^{L_p}(w) = w^\sigma (L_p)^{1-\alpha(\sigma-1)} - \bar{A}_p^{\sigma-1}P_p^{\sigma-1} w L_p  - \sum_{j = 1}^{n^*} T_{pj}^{1-\sigma} \bar{A}_p^{\sigma-1} P_j^{\sigma-1} w_j L^{-}_j.
\end{equation*}

We are interested in the properties of the limit of $w_p(L_p)$ as $L_p \to 0^+$. We then distinguish different cases. 

\textbf{Case 1: $\alpha > 1/(\sigma - 1)$.} First note that for any $w >0$, we have that $\lim_{L_p\to 0^+}f^{L_p}(w) = +\infty$. At the same time $f^{L_p}(0) < 0$ for all $L_p >0$. Then, arguing by contradiction, it is easy to show that $w^*_p = \lim_{L_p\to 0^+}w_p(L_p)=0$. Indeed, if $\lim_{L_p\to 0^+}w_p(L_p) = w^*_p >0$ we would have that $\lim_{L_p\to 0^+}f^{L_p}(w^*_p) = +\infty \neq 0 = \lim_{L_p\to 0^+}f^{L_p}(w_p (L_p) )$.

\textbf{Case 2: $\alpha = 1/(\sigma - 1)$.} Here, for $w \geq 0$, we have 
$$\lim_{L_p\to 0^+}f^{L_p}(w) = w^\sigma - \sum_{j = 1}^{n^*} T_{pj}^{1-\sigma} \bar{A}_p^{\sigma-1} P_j^{\sigma-1} w_j L^{-}_j$$ 
and it is easy to show that the convergence is uniform on all compact subsets of $[0,+\infty)$.
 We then have that $w^*_p$ satisfies
\begin{equation}\label{eq:wstarp}
(w^*_p )^{\sigma} = \sum_{i = 1}^{n^*} T_{pj}^{1-\sigma}\bar{A}_p^{(\sigma - 1)} P_j^{\sigma - 1} w_j L_j.
\end{equation}

\textbf{Case 3: $\alpha < 1/(\sigma - 1)$.}
In this case,
for $w \geq 0$ we have 
$$\lim_{L_p\to 0^+}f^{L_p}(w) = - \sum_{j = 1}^{n^*} T_{pj}^{1-\sigma} \bar{A}_p^{\sigma-1} P_j^{\sigma-1} w_j L^{-}_j <0$$ 
and the convergence is uniform on all compact subsets of $[0,+\infty)$. Then, arguing by contradiction, it is easy to show that $w^*_p = +\infty$. Indeed, if $w^*_p <+\infty$, we would have $\lim_{L_p\to 0^+}f^{L_p}(w^*_p) <0 \neq 0 =\lim_{L_p\to 0^+}f^{L_p}(w_p(L_p))$. \bigskip

Now, consider the potential weight $\lambda_p = \frac{1}{\delta} \log \frac{w^*_p}{P_p}$, where $P_p$ is defined by equation $\eqref{eq:P}$. Given the results above, we immediately get 
\begin{equation*}
    \lambda_p = 
    \begin{cases}
        - \infty, \quad \alpha > \frac{1}{\sigma - 1}, \\ 
        + \infty, \quad \alpha < \frac{1}{\sigma - 1}.
    \end{cases}
\end{equation*}
For $\alpha = 1/(\sigma-1)$, both $w^*_p$ and $P_p$ are well-defined, and we can compute $\lambda_p$ directly. At an equilibrium with $n^*$ cities, welfare equalization implies
\begin{equation*}    
P_i = \frac{1}{V} B_i(\lambda) w_i L_i^{\beta}
\end{equation*}
for all $y_i \in \Y^*$. Replacing the price index in \eqref{eq:wstarp}, we get
\begin{equation*}
(w^*_p)^{\sigma} = V^{1-\sigma} \sum_{i = 1}^{n^*} T_{pj}^{1-\sigma}\bar{A}_p^{(\sigma - 1)} B_j(\lambda)^{\sigma - 1} w_j^{\sigma} L_j^{1+(\sigma-1)\beta}.   
\end{equation*}
Furthermore, with $\alpha = 1/(\sigma - 1)$, equation \eqref{eq:wscale} becomes:
\begin{equation*}    
 w_i = c  \bar{A}_i^{\sigmatil \sigma} B_i(\lambda)^{-\sigmatil \sigma} L_i^{-\sigmatil\beta }
\end{equation*}
for all $y_i \in \Y^*$, where $c > 0$ is a positive constant. Replacing the wage in the expressions for $w^*_p$ and $P_p$, we get
\begin{align*}
    w_p^{\sigma} &= c^{\sigma} V^{1-\sigma} \sum_{j = 1}^{n^*} T_{pj}^{1-\sigma} \bar{A}_p^{\sigma-1} \bar{A}_j^{\tilde{\sigma}\sigma}  B_j(\lambda)^{\tilde{\sigma}(\sigma-1)} L_j^{\sigmatil \gamma_2} \\
    P_p^{1-\sigma}  &= c^{1-\sigma} \sum_{j = 1}^{n^*} T_{pj}^{1-\sigma}  \bar{A}_j^{\tilde{\sigma}\sigma}  B_j(\lambda)^{\tilde{\sigma}(\sigma-1)} L_j^{\sigmatil \gamma_2},
\end{align*}
with $\gamma_2 = 1/\sigmatil + (\sigma - 1) \beta$. Therefore, 
\begin{align*}
    \frac{w^*_p}{P_p} = e^{\delta \lambda_p} &=  V^{\frac{1-\sigma}{\sigma} } \bar{A}_p^{\frac{\sigma-1}{\sigma} } \left(\sum_{j = 1}^{n^*} T_{pj}^{1-\sigma}  \bar{A}_j^{\tilde{\sigma}\sigma}  B_j(\lambda)^{\tilde{\sigma}(\sigma-1)} L_j^{\sigmatil \gamma_2} \right)^{\frac{1}{\sigmatil \sigma}} \iff \\
    e^{\delta \sigmatil \sigma \lambda_p} &=  V^{\frac{1}{\beta}} \sum_{j = 1}^{n^*} T_{pj}^{1-\sigma} \bar{A}_p^{\sigmatil(\sigma-1)}  \bar{A}_j^{\tilde{\sigma}\sigma}  B_j(\lambda)^{-\frac{1}{\beta}} e^{ - \frac{\delta}{\beta} \sigmatil \gamma_2
 \lambda_j }, 
\end{align*}
where we have used again the welfare equalization condition to express the equilibrium labor force in terms of the Voronoi weight for the business districts in $\Y^*$. This expression pins down $\lambda_p$ for the case $\alpha = 1/(\sigma - 1)$. Next, we observe that in the case $\alpha = 1 / (\sigma - 1)$ equation \eqref{eq:lambda_eq} also takes the form: 
\begin{equation*}
    	e^{\delta \sigmatil \sigma \lambda_i } = V^{\frac{1}{\beta}} \sum_{j = 1}^{n^*} T_{ij}^{1-\sigma} \bar{A}_i^{\tilde{\sigma}(\sigma-1)} \bar{A}_j^{\tilde{\sigma}\sigma} B_j(\lambda)^{-\frac{1}{\beta}} e^{ - \frac{\delta}{\beta} \sigmatil \gamma_2 \lambda_j}. \qedhere
\end{equation*}
\end{proof}

\begin{remark}\label{remark:lamdap}
When $\alpha \geq 1/(\sigma - 1)$, such that $w^*_p$ is finite, it is clear that total income in $y_p $, $w_p L_p$, goes to zero as $L_p$ approaches zero. When $\alpha < 1/(\sigma - 1)$, such that $w^*_p = + \infty$, $w_p L_p$ goes to zero as $L_p$ approaches zero as long as $\alpha \geq -1$. To see this, repeat the proof of Proposition \ref{prop:lambdap} with the change of variables $(wL)_p = w_p L_p$. The restriction $\alpha \geq - 1$ implies that output does not decrease with the size of the labor input, and does not go to infinity as the local labor force approaches zero. 

\end{remark}

\subsection{Proof of Theorem \ref{theo:theo1}}\label{app:proofstheo1}

The proof is based on the application of the Schauder and Banach fixed point theorems and will be presented at the end of this section. Before that, we need to introduce some notation and four technical lemmas.

Because $\gamma_1 \neq 0$ and $V > 0$, we can introduce the change of variables 
\begin{equation*}
\tilde{\lambda}_i = -\frac{\delta}{\beta} \tilde{\sigma}\gamma_{1} \left(\lambda_i - \frac{\alpha}{\alpha + \beta} \frac{1}{\delta} \log V \right) , \quad i=1, \dots, n^*. 
\end{equation*}
For all $i,j = 1, \dots, n^*$, define $K_{ij} = T_{ij}^{(1-\sigma)} \bar{A}_i^{\tilde{\sigma}(\sigma-1)} \bar{A}_j^{\tilde{\sigma}\sigma}$ and $\tilde{B}_i$ such that $\tilde{B}_i(\tilde{\lambda}) = B_i(\lambda)$. After taking logarithms, the equilibrium system \eqref{eq:lambda_eq} in the transformed variables becomes
\begin{equation*}
	\lambdatil_i = \log \sum_{j = 1}^{n^*} K_{ij} \Btil_i(\lambdatil)^{\sigmatil \phi_1} \Btil_j(\lambdatil)^{\sigmatil \phi_2} e^{ \frac{\gamma_2}{\gamma_1} \lambdatil_j }, \quad i = 1, \dots, n^*.
	\end{equation*}
For $\delta >0$, define the unbounded open set
\begin{equation*}
\begin{split}
\tilde \Lambda = \left\{ \lambdatil \in \R^{n^*}: 
\frac{\delta}{\beta} \sigmatil \gamma_1 d_j(y_i) < \lambdatil_i - \lambdatil_j < - \frac{\delta}{\beta} \sigmatil \gamma_1 d_i(y_j),  \; i,j = 1,\dots, n^*,  i \neq j \right\},
\end{split}
\end{equation*}
such that, for $\gamma_1 > 0$, $\lambda \in \Lambda$ if and only if $\lambdatil \in \tilde \Lambda$, with the inequalities reversed for $\gamma_1 < 0$. For $\delta > 0$ and $k \in (0,1)$, define the unbounded closed set
\begin{equation}
\begin{split}\label{def:gammak}
\tilde \Lambda^k = \left\{ \lambdatil \in \R^{n^*}:
\frac{\delta}{\beta} \sigmatil \gamma_1 kd_j(y_i) \leq \lambdatil_i - \lambdatil_j \leq - \frac{\delta}{\beta} \sigmatil \gamma_1 kd_i(y_j), \; i,j = 1,\dots, n^*,  i \neq j \right\}.
\end{split}
\end{equation}
Note that $\tilde \Lambda^k\subset \tilde \Lambda$ for every $k\in(0,1)$, and that both $\tilde \Lambda$ and $\tilde \Lambda^k$ depend on $\delta$. Consider the map $g \colon\R^{n*}\to \R^{n*}$,
\begin{equation*}
g_i(\tilde {\lambda}) = \log \sum_{j = 1}^{n^*} K_{ij} \Btil_i(\lambdatil)^{\sigmatil \phi_1} \Btil_j(\lambdatil)^{\sigmatil \phi_2} e^{ \frac{\gamma_2}{\gamma_1} \lambdatil_j }, \quad i =1,\dots,n^*.
\end{equation*}
We study the fixed points of $g$ in $\tilde \Lambda^k$, since $\cup_{k'\in (0,1)} \tilde \Lambda^{k'} = \tilde \Lambda$.

\begin{lemma}\label{lemma:Bbound}
Let $k \in (0,1)$. There exists $\delta_0 >0$ such that, for $\delta \geq \delta_0$, we have
$$
\frac{c^{\prime}}{\delta^{-2\beta}} \bar b_{min} \leq \tilde B_i(\tilde \lambda) \leq \frac{C^{\prime}}{\delta^{-2\beta}} \bar b_{max}, \qquad \lambdatil \in \tilde \Lambda^k,
$$
for all $i = 1,\dots,n^*$, and for some constants $c^{\prime},C^{\prime} >0$, depending only on $k$.
\end{lemma}
\begin{proof}
Let $\delta >0$ and, for $\lambdatil \in \tilde \Lambda^k$, consider the associated Voronoi tessellation $\{ \Omega_i\}_{i=1}^{n^*}$. 
Let $B_\varepsilon(y_i)$ denote an open ball centered in $y_i$ of radius $\varepsilon$ in the Euclidean distance. 
Because $\lambdatil \in \tilde \Lambda^k$, we can choose $\epsilon$ small enough to ensure that $B_\varepsilon (y_i) \subset \Omega_i$, independently of $\delta$. 
Because $\beta < 0$ and $\bar b$ is bounded (Assumption \ref{ass:Ab}), \begin{align*}
\bar b_{min} \left( \int_{B_\varepsilon(y_i)}\frac{1}{e^{-\frac{\delta}{\beta}d_i(x)}}dx \right)^{-\beta} &\leq  \Btil_i(\lambdatil) \\ 
& \leq  \bar b_{max} \left( \int_{B_\varepsilon(y_i)}\frac{1}{e^{-\frac{\delta}{\beta}d_i(x)}}dx  + \int_{\Omega_i \setminus B_\varepsilon(y_i)}\frac{1}{e^{-\frac{\delta}{\beta}d_i(x)}}dx\right)^{-\beta}.
\end{align*}
Thanks to Assumption \ref{ass:doublelip}, we have that
\begin{align*}
\bar b_{min} \left( \int_{B_\varepsilon(y_i)}\frac{1}{e^{-\frac{\delta}{\beta}C\|x-y_i\|}}dx \right)^{-\beta} &\leq  \Btil_i(\lambdatil) \\ 
& \leq  \bar b_{max} \left( \int_{B_\varepsilon(y_i)}\frac{1}{e^{-\frac{\delta}{\beta}c\|x-y_i\|}}dx  + \frac{|\Omega_i\setminus B_\varepsilon(y_i)|}{e^{-\frac{\delta}{\beta} c\epsilon}}\right)^{-\beta}.
\end{align*}
for all $x \in B_\varepsilon(y_i)$, with constants $c, C >0$ possibly depending on $\varepsilon$.
By passing in polar coordinates and integrating by parts, we can solve explicitly the following integral:
\begin{align*}
\int_{B_\varepsilon(y_i)}\frac{1}{e^{-\frac{\delta}{\beta}\|x-y_i\|}}dx = 2\pi \int_0^\varepsilon e^{\frac{\delta}{\beta}r}rdr = 2 \pi \left[ \left( 1 - e^{\frac{\delta \varepsilon}{\beta}} \right) \frac{ \beta^2}{\delta^2}+\varepsilon  e^{\frac{\delta \varepsilon}{\beta}} \frac{\beta}{\delta} \right].
\end{align*}
As $\delta \to +\infty$, and given that $\beta < 0$, the term $\delta^{-2}$ dominates the decay, while the other terms decrease exponentially. Then, there exists $\delta_0 >0$ such that for $\delta \geq \delta_0$ we have
$$
\frac{c'}{\delta^{-2\beta}} \bar b_{min} \leq \tilde B_i(\tilde \lambda) \leq \frac{C'}{\delta^{-2\beta}} \bar b_{max},
$$
for some constants $c^{\prime},C^{\prime}$, such that $0 < c^{\prime} < C^{\prime}$.
\end{proof}

\begin{lemma}\label{lemma:gmap}
Let $k\in (0,1)$. There exists $\delta_1$ such that, for $\delta > \delta_1$, $g$ maps $\tilde \Lambda^k$ onto itself. 
\end{lemma}
\begin{proof}
Choose $\lambdatil \in \tilde \Lambda^k$. First, consider the case $\gamma_1 > 0$. The goal is to show that, under the assumptions of the lemma, 
\begin{equation}\label{eq:gammabounds}
\frac{\delta}{\beta} \sigmatil \gamma_1 k d_j(y_i) \leq g_i(\lambdatil) - g_j(\lambdatil)
\leq -\frac{\delta}{\beta} \sigmatil \gamma_1 kd_i(y_j),
\end{equation}
for all $i,j = 1, \dots, n^*$, $i\neq j$, where
\begin{align*}
g_i(\lambdatil) - g_j(\lambdatil) = \sigmatil (\sigma - 1) \log \frac{\bar A_i}{\bar A_j} + \sigmatil \phi_1 \log \frac{ \Btil_i(\lambdatil)}{\Btil_j(\lambdatil)} + \log \frac{ \sum_{k = 1}^{n^*}  T_{ik}^{1-\sigma} \bar{A}_k^{\sigmatil \sigma}\Btil_k(\lambdatil)^{\sigmatil \phi_2} e^{ \frac{\gamma_2}{\gamma_1} \lambdatil_k } }{
\sum_{k = 1}^{n^*}  T_{jk}^{1-\sigma} \bar{A}_k^{\sigmatil \sigma}  \Btil_k(\lambdatil)^{\sigmatil \phi_2} e^{ \frac{\gamma_2}{\gamma_1} \lambdatil_k }}.
\end{align*}
Due to Assumption \ref{ass:Ab}, $\bar A_i / \bar A_j$ is bounded above and below. Further, due to Assumption \ref{ass:T} $\max_{y_k \in \Y^*} T_{jk}$ is well-defined, and $\min_{y_k \in \Y^*} T_{ik} = 1$. Thus, we have that
\begin{equation*}
    g_i(\lambdatil) - g_j(\lambdatil) \leq \sigmatil (\sigma - 1) \log \frac{\bar A_i}{\bar A_j} + \sigmatil \phi_1 \log \frac{ \Btil_i(\lambdatil)}{\Btil_j(\lambdatil)} + (\sigma - 1) \log \max_{y_k \in \Y^*} T_{jk}.
\end{equation*}
Next, choose $\delta > \delta_0$, with $\delta_0$ given in Lemma \ref{lemma:Bbound}, such that the bounds on $\Btil$ given in that Lemma apply. Then, we have that
\begin{align*}
g_i(\lambdatil) - g_j(\lambdatil) \leq 
 \sigmatil (\sigma - 1) \log \frac{\bar A_i}{\bar A_j}  +  c + \sigmatil | \phi_1 | \log \left( \frac{\bar b_{max}}{\bar b_{min}} \right) + (\sigma - 1) \log \max_{y_k \in \Y^*} T_{jk},
\end{align*}
where $c> 0$ is constant not depending of $\delta$. Because the right-hand side of this inequality does not depend on $\delta$, we can always find $\delta_1 \geq \delta_0$ such that for all $\delta > \delta_1$
\begin{align}\label{eq:suffcondexist}
c + \sigmatil (\sigma - 1) \log \frac{\bar A_i}{\bar A_j}  +  \sigmatil | \phi_1 | \log \left( \frac{\bar b_{max}}{\bar b_{min}} \right) + (\sigma - 1) \log \max_{y_k \in \Y^*} T_{jk} \leq -\frac{\delta}{\beta} \sigmatil \gamma_1 k d_i(y_j), 
\end{align}
holds for all $i,j = 1, \dots, n^*$, $i\neq j$. For any such $\delta$, all the inequalities in \eqref{eq:gammabounds} are satisfied. The same is true when $\gamma_1 <0 $, with the direction of all inequalities reversed.
\end{proof}

\begin{remark}\label{remark:suffcondexist}
If we also invoke Assumption \ref{ass:T2}, then Assumption \ref{ass:trineq} (triangle inequality) implies that, for any $y_i, y_j, y_k \in \Y$, $ T_{jk} \leq  T_{ij}  T_{ik}$. Therefore, we can derive a sharper bound for differences in trade market access across business districts,
\begin{align}\label{eq:suffcondexistT2}
\log \frac{ \sum_{k = 1}^{n^*}  T_{ik}^{1-\sigma} \bar{A}_k^{\sigmatil \sigma}\Btil_k(\lambdatil)^{\sigmatil \phi_2} e^{ \frac{\gamma_2}{\gamma_1} \lambdatil_k } }{
\sum_{k = 1}^{n^*}  T_{jk}^{1-\sigma} \bar{A}_k^{\sigmatil \sigma}  \Btil_k(\lambdatil)^{\sigmatil \phi_2} e^{ \frac{\gamma_2}{\gamma_1} \lambdatil_k }} \leq (\sigma - 1) \log T_{ij} = \tau(\sigma - 1) d_i(y_j).
\end{align}
As a result, we can write the sufficient condition for the existence of a spatial equilibrium in $\Y^*$ in Lemma \ref{lemma:gmap}, for the case $\gamma_1 > 0$, as 
\begin{align*}
         c + \sigmatil (\sigma - 1) \log \frac{\bar A_i}{\bar A_j}  +  \sigmatil | \phi_1 | \log \left( \frac{\bar b_{max}}{\bar b_{min}} \right) < \left( -\frac{\delta}{\beta} \sigmatil \gamma_1 - \tau(\sigma - 1) \right) d_i(y_j),
\end{align*}
for all $y_i, y_j \in \Y^{*}$ with $y_i \neq y_j$. These inequalities will never hold for all pairs of business districts unless $-(\delta/\beta) \sigmatil \gamma_1 > \tau(\sigma - 1)$.
\end{remark}

\begin{lemma}\label{lemma:gcontraction}
Let $k\in (0,1)$. We have that 
\begin{equation*}
\max_{i = 1, \dots, n^*}  \max_{\lambdatil \in \tilde \Lambda^k} \sum_{k=1}^{n^*}\left|\frac{\partial g_i}{\partial  \lambdatil_k}(\lambdatil)\right| \leq \left\vert \frac{\gamma_2}{\gamma_1} \right\vert +  \bigg(  2 (n^* - 1) \vert \phi_1 \vert  + (2 n^* - 1) \vert \phi_2 \vert  \bigg) \tilde \eta,
\end{equation*}
where $\tilde \eta = \max_{i\neq k}  \sup_{\lambdatil \in \tilde  \Lambda^k } \left\vert \frac{ \partial \Btil_i(\lambdatil) }{ \partial \lambdatil_k } \frac{1}{\Btil_i(\lambdatil) } \right\vert $.
\end{lemma}
\begin{proof}
For all $i,j =1,\dots,n^*$, let 
\begin{equation}\notag
f_{ij}(\tilde{\lambda}) = K_{ij} \tilde{B}_i( \tilde{\lambda} )^{\sigmatil \phi_1} \tilde{B}_j(\tilde{\lambda})^{\sigmatil \phi_2}e^{\frac{\gamma_2}{\gamma_1}\tilde \lambda_j},
\end{equation}
so that $g_i(\tilde \lambda) = \log \sum_{j=1}^{n^*} f_{ij}(\tilde \lambda)$, and let
\begin{equation}\notag
\tilde \eta_{ij}(\lambdatil) = \frac{ \partial \Btil_i(\lambdatil) }{ \partial \tilde\lambda_j } \frac{1}{\Btil_i(\lambdatil)} .
\end{equation} 
Henceforth, we omit the dependency on $\tilde \lambda$. We have 
\begin{align*}
\sum_{k = 1}^{n^*} \left| \frac{\partial g_i}{\partial \tilde{\lambda}_k} \right| = \sum_{k = 1}^{n^*} \left| \frac{ \sum_{j=1}^{n^*} \partial f_{ij} / \partial \tilde{\lambda}_k}{\sum_{j=1}^{n^{*}}  f_{ij}} \right|  \leq \sum_{k = 1}^{n^*}  \frac{ \sum_{j=1}^{n^*} \left| \partial f_{ij} / \partial \tilde{\lambda}_k \right| }{\sum_{j=1}^{n^{*}}  f_{ij}}.
\end{align*}
For all $j,k = 1, \dots, n^*$, we have 
\begin{align*}
\frac{\partial f_{ik}}{\partial \tilde{\lambda}_k} &=
\left( \frac{\gamma_2}{\gamma_1} + \sigmatil \phi_1 \tilde \eta_{ik} +  \sigmatil \phi_2 \tilde \eta_{kk} \right) f_{ik}, \\
\frac{\partial f_{ij}}{\partial \tilde{\lambda}_k} &=
\left(  \sigmatil \phi_1 \tilde \eta_{ik} +  \sigmatil \phi_2 \tilde \eta_{jk} \right) f_{ij}, \quad j \neq k.
\end{align*}
Using these derivatives into the previous inequality, we obtain
\begin{equation*}
    \sum_{k =1}^{n^*} \left| \frac{\partial g_i}{\partial \tilde{\lambda}_k} \right| \leq \left|\frac{\gamma_2}{\gamma_1}\right| + \sigmatil |\phi_1 | \sum_{k =1}^{n^*} \left| \tilde \eta_{ik} \right|  + \sigmatil |\phi_2|     
      \sum_{k =1}^{n^*} \sum_{\ell = 1 }^{n^{*}} \frac{f_{i\ell}}{\sum_{j=1}^{n^*}f_{ij}} \left|\tilde \eta_{\ell k}\right|.
\end{equation*}
Next, we use some basic properties of the shape derivative (see Appendix \ref{app:shapeder}). First, we have  $\sum_{k= 1}^{n^*} \left| \eta_{ik} \right| = \eta_{ii} + \sum_{k \in \NN_i} \left| \eta_{ik} \right| $, because $\eta_{ik}$ is zero unless either $i = k$ or  $y_k$ is $y_i$'s neighbor. Using this fact, we can rewrite the previous inequality as
\begin{align*}
    \sum_{k =1}^{n^*} \left| \frac{\partial g_i}{\partial \tilde{\lambda}_k} \right| &\leq \left|\frac{\gamma_2}{\gamma_1}\right| \\& + \sigmatil |\phi_1 | \left(\tilde \eta_{ii} + \sum_{k \in \NN_i}^{n^*} \left| \tilde \eta_{ik} \right| \right)  \\& + \sigmatil |\phi_2|   \sum_{k =1}^{n^*}  \left( 
 \frac{ f_{ik}  }{\sum_{j=1}^{n^*}f_{ij}} \tilde \eta_{kk} + \sum_{\ell \in \NN_k} \frac{f_{i\ell} } {\sum_{j=1}^{n^*}f_{ij}} |\tilde \eta_{\ell k} |
     \right)  
\end{align*}
Second, we have $\tilde \eta_{ii} = \sum_{k \in \NN_i} \vert \tilde \eta_{ik} \vert$, as the own-weight effect is the sum of the cross-weight effects acting, with a negative sign, on the opposite side of each border segment. Because $|\eta_{ik}| \leq \eta$ for $i \neq k$, and the number of neighbors is at most $n^*-1$, we find
\begin{equation}\notag
\sum_{k = 1}^{n^*} \left\vert \frac{\partial g_i(\tilde{\lambda})}{\partial \tilde{\lambda}_k} \right \vert \leq \left\vert \frac{\gamma_2}{\gamma_1} \right\vert +  \sigmatil \bigg(  2 (n^* - 1) \vert \phi_1 \vert  + (2 n^* - 1) \vert \phi_2 \vert  \bigg) \tilde \eta.
\end{equation}
Because the right-hand side is independent both of $i$ and $\lambdatil$, the same inequality holds for the maximum with respect to $i$ in $\{1, \dots, n^*\}$ and the supremum with respect to $\lambdatil$ inside the closed set $\tilde \Lambda^k$. 
\end{proof}
\begin{lemma}\label{lemma:etabound}
We have that 
$\tilde \eta = \max_{i\neq k}  \max_{\lambdatil \in \tilde  \Lambda^k } \left\vert \frac{ \partial \Btil_i(\lambdatil) }{ \partial \lambdatil_k } \frac{1}{\Btil_i(\lambdatil) } \right\vert \to 0$ as $\delta \to +\infty$.
\end{lemma}

\begin{proof}

    Let $\tilde \eta_{ik} = \frac{\partial \tilde B_i}{\partial \tilde \lambda_k} \frac{1}{ \tilde B_i}$, for $i\neq k$. Then,
\begin{align*}
\tilde \eta_{ik}  = -\frac{\beta}{\delta \tilde\sigma \gamma_1} \frac{\partial B_i}{\partial \lambda_k}\frac{1}{ B_i}= \frac{\beta^2}{\delta \tilde\sigma \gamma_1} \frac{\int_{\Gamma_{ik}} \left(\frac{\bar b(x)}{D(x,y_i)}\right)^{-\frac{1}{\beta}} \frac{\partial \omega_{ik}}{\partial \lambda_k }(x, \lambda) \cdot \nu(x) \d \sigma(x) }{\int_{\Omega_i} \left(\frac{\bar b(x)}{D(x,y_i)}\right)^{-\frac{1}{\beta}} \d x}.
\end{align*}
In the first identity we have applied the chain rule, while in the second one we have applied formula \eqref{eq:shapeder} for the shape derivative. First, we bound the denominator from below. By Lemma \ref{lemma:Bbound}, we can always find $\delta_0 > 0$ such that, for $\delta > \delta_0$, we have
\begin{equation*}
  \int_{\Omega_i} \left(\frac{\bar b(x)}{D(x,y_i)}\right)^{-\frac{1}{\beta}} \d x  = B_i(\lambda)^{-\frac{1}{\beta}} = \Btil_i(\lambdatil)^{-\frac{1}{\beta}}  \geq  \frac{c^{\prime \prime}}{\delta^2} \bar b_{min}^{-\frac{1}{\beta}},
\end{equation*}
for some constant $c^{\prime \prime} > 0$ independent of $\delta$. In the proof of the Lemma, we showed that there exists $\varepsilon>0$ such that $B_\varepsilon(y_i)$ is contained in $\Omega_i$, with $\varepsilon$ independent of $\delta$ and $i=1,\dots,n^*$. Second, we bound the numerator from above. We have: 
\begin{align*}
\left|\int_{\Gamma_{ik}} \left(\frac{\bar b(x)}{D(x,y_i)}\right)^{-\frac{1}{\beta}} \frac{\partial \omega_{ik}}{\partial \lambda_k }(x, \lambda) \cdot \nu(x) \d \sigma(x)\right|  &\leq \int_{\Gamma_{ik}}  \left(\frac{\bar b(x)}{D(x,y_i)}\right)^{-\frac{1}{\beta}} \left| \frac{\partial \omega_{ik}}{\partial \lambda_k }(x, \lambda) \cdot \nu(x) \right| \d \sigma(x) \notag \\ 
& \leq \bar b_{max}^{-\frac{1}{\beta}} K e^{\frac{\delta}{\beta} \epsilon },
\end{align*}
where we used the fact that $\bar b(x)/D(x,y_i)$ is a positive function, the a priori bounds on $\bar b(x)$, the Cauchy-Schwarz inequality, the fact that the distance of $\Gamma_{ik}$ to $y_i$ is greater than $\varepsilon$ and the fact that  $\int_{\Gamma_{ik}}  \left| \frac{\partial \omega_{ik}}{\partial \lambda_k }(x, \lambda) \cdot \nu(x) \right| \d \sigma(x) \leq K$ for some $K > 0$ and all $y_i, y_k \in \Y^{*}$, because $\omega$ is a parametrization of the Voronoi region's boundary with Lipschitz distance functions. 
By combining these results, we find
\begin{equation}\notag
\left| \frac{\int_{\Gamma_{ik}} \left(\frac{\bar b(x)}{D(x,y_i)}\right)^{-\frac{1}{\beta}} \frac{\partial \omega_{ik}}{\partial \lambda_k }(x, \lambda) \cdot \nu(x) \d \sigma(x) }{\int_{\Omega_i} \left(\frac{\bar b(x)}{D(x,y_i)}\right)^{-\frac{1}{\beta}} \d x} \right| \leq 
 \frac{ K }{c^{\prime \prime}} \left( \frac{\bar b_{max}}{\bar b_{min}} \right)^{-\frac{1}{\beta}}\delta^2  e^{\frac{\delta}{\beta} \epsilon}.
\end{equation}
Since $\beta < 0$, the right-hand side approaches zero as $\delta \to +\infty$. Because the right-hand side is independent both of $i$ and $\lambdatil$, the same inequality holds for the maximum with respect to $i$ in $\{1, \dots, n^*\}$ and $\lambdatil$ in the closed set $\tilde \Lambda^k$.
\end{proof}
\begin{remark}
Let $x(t)$ be a point on the boundary $\Gamma_{ik}$ at ``time'' $t$, where we perturb $\lambda_k$ by $t$. Then,
\[d(x(t),y_i) - \lambda_i = d(x(t),y_k) - (\lambda_k + t).\]
Differentiating both sides with respect to $t$ at $t = 0$, we obtain
\[\nabla d(x,y_i) \cdot \frac{\d x}{\d t} = \nabla d(x,y_k) \cdot \frac{\d x}{\d t} - 1\]
Here, $\nabla d(x,y_i)$ and $\nabla d(x,y_k)$ are vectors pointing from $y_i$ and $y_k$ to $x$, respectively. The vector $\d x/\d t$ corresponds to $\partial \omega_{ik}/\partial \lambda_k$. The normal vector $\nu(x)$ at $x$ is parallel to the difference between these gradients:
\[\nu(x) = \frac{\nabla d(x,y_i) - \nabla d(x,y_k)}{\|\nabla d(x,y_i) - \nabla d(x,y_k)\|}\]
Therefore,
\[\frac{\partial \omega_{ik}}{\partial \lambda_k}(x, \lambda) \cdot \nu(x) = -\frac{1}{\|\nabla d(x,y_i) - \nabla d(x,y_k)\|}.\]
\end{remark}
Given these preliminary results, we can now present the proof of Theorem \ref{theo:theo1}. 
\begin{proof}[Proof of Theorem \ref{theo:theo1}]

We will study the fixed points of $g$ inside $\tilde \Lambda^k$, since $\cup_{k'\in (0,1)} \tilde \Lambda^{k'} = \tilde \Lambda$. If we find a solution $\lambdatil^* = g(\lambdatil^*)$, we can then use the aggregate population constraint \eqref{eq:aggpop} and the welfare equalization condition \eqref{eq:weq} to compute the welfare scalar $V$, as follows:
\begin{align*}
     L &= \sum_{i = 1}^{n^*} L_i = V^{\frac{1}{\beta}} \sum_{i = 1}^{n^*} B_i(\lambda)^{-\frac{1}{\beta}} e^{-\frac{\delta}{\beta}\lambda_i}  \iff \\
    V & = \left(\sum_{i = 1}^{n^*} B_i(\lambda)^{-\frac{1}{\beta}} e^{-\frac{\delta}{\beta}\lambda_i} \right)^{-\beta} L^{\beta} \iff \\ 
    V & = \left(\sum_{i = 1}^{n^*} \tilde B_i(\lambdatil)^{-\frac{1}{\beta}} e^{\frac{1}{\sigmatil \gamma_1}\lambdatil_i} \right)^{-(\alpha + \beta) } L^{\alpha + \beta}.
\end{align*}
Then, we can back out the Voronoi weights in the original variables,
\begin{equation*}
    \lambda^* = -\frac{\beta}{\delta}\frac{1}{\sigmatil \gamma_1} \lambdatil^*_i + \frac{\alpha}{\alpha + \beta} \frac{1}{\delta} \log V.
\end{equation*}
The welfare scalar $V$ and the vector of Voronoi weights $\lambda^*$ are a solution to the equilibrium system of equations \eqref{eq:lambda_eq}. From the other equilibrium conditions, we can then uniquely recover the other components of a $\Y^*$-centric equilibrium. 

\textit{Proof of \ref{cond:1a}.} Fix a coordinate $y_{i_0} \in \Y^*$, and define 
\begin{equation*}
\hat \lambda_i  = \lambdatil_i - \lambdatil_{i_0} = -\frac{\delta}{\beta} \sigmatil \gamma_1 (\lambda_i - \lambda_{i_0}), \quad i=1, \dots, n^*, i \neq i_0, 
\end{equation*}
with $\hat \lambda_{i_0} = 0$. Accordingly,  for all $i = 1, \dots, n^*$ define $\hat B_i$ such that $\hat B_i (\hat \lambda) = B_i(\lambda)$. Consider the map $\hat{g}: \R^{n^*-1} \to \R^{n^*-1}$, 
\begin{equation*}
\hat{g}_i(\hat \lambda) =  \log \frac{ \sum_{k = 1}^{n^*} K_{ik} \hat B_i(\hat\lambda)^{\sigmatil \phi_1} \hat B_k(\hat\lambda)^{\sigmatil \phi_2} e^{ \frac{\gamma_2}{\gamma_1} \hat\lambda_k } }{
	\sum_{k = 1}^{n^*} K_{i_0 k} \hat B_{i_0 }(\hat\lambda)^{\sigmatil \phi_1} \hat B_k(\hat\lambda)^{\sigmatil \phi_2} e^{ \frac{\gamma_2}{\gamma_1} \hat\lambda_k }}, \quad i = 1, \dots, n^*-1. 
\end{equation*}
Finally, define the bounded closed set
\begin{equation*}
\begin{split}
\hat \Lambda^k = \bigg\{ \hat \lambda \in \R^{n^* -1}: & \\
\frac{\delta}{\beta} \sigmatil \gamma_1 k d_{i_0}(y_{i}) \leq & \; \hat \lambda_i  \leq - \frac{\delta}{\beta} \sigmatil \gamma_1 k d_{i}(y_{i_0}), \; i = 1,\dots, n^*, i \neq i_0 ,
\\
\frac{\delta}{\beta} \sigmatil \gamma_1 k d_j(y_i) \leq \hat \lambda_i &- \hat \lambda_j \leq - \frac{\delta}{\beta} \sigmatil \gamma_1 k d_i(y_j),  \; i,j = 1,\dots, n^*, i,j \neq i_0,  i \neq j \bigg\}.
\end{split}
\end{equation*}
Clearly, for $\gamma_1 > 0$, $\lambdatil \in \tilde \Lambda^k \iff \hat \lambda \in \hat \Lambda^k$, with the inequalities reversed for $\gamma_1 < 0$. 

By construction, $\hat \Lambda^k$ is a nonempty compact subset of $\R^{n^*-1}$. It is easy to see that $\hat \Lambda^k$ is also convex. Now, choose $\delta > \delta_1$, with $\delta_1 > 0$ given in 
Lemma \ref{lemma:gmap}. Because $\hat{g}_{i}(\hat{\lambda})= g_i(\lambdatil) - g_{i_0}(\lambdatil)$ for all $i \neq i_0$ and $\hat{g}_{i}(\hat{\lambda}) - \hat{g}_{j}(\hat{\lambda}) = g_i(\lambdatil) - g_{j}(\lambdatil)$ for all $i\neq j$ and $i,j \neq i_0$, Lemma \ref{lemma:gmap} implies that $\hat{g}$ maps $\hat \Lambda^k$ onto itself. Therefore, $\hat{g}$ is a continuous function that maps a nonempty convex compact subset of $\R^{n^*-1}$ onto itself. By Schauder fixed point theorem, it has a fixed point in $\hat \Lambda^k$. Let $\hat \lambda^* \in \hat \Lambda^k$ be one such fixed point. Then $\tilde \lambda^*$ defined as $\tilde \lambda^*_i = \hat \lambda^*_i$ for $i\neq i_0$ and $\tilde \lambda^*_{i_0} =0$ is a fixed point of $g$.

\textit{Proof of  \ref{cond:1b}.} 
The equilibrium system \eqref{eq:lambda_eq} is a version of the system of equations studied in \cite{allen2024}. Choose $\delta > \delta_1$, with $\delta_1 > 0$ given in Lemma \ref{lemma:gmap}, such that $g:\tilde \Lambda^k \to \tilde \Lambda^k$. In Lemma \ref{lemma:gcontraction}, we show that 
\begin{equation}\notag
   \max_{i = 1, \dots, n^*}  \max_{\lambdatil \in \tilde \Lambda^k} \left( \sum_{k =1 }^{n^*} \left\vert  \frac{\partial g_i}{\partial \tilde{\lambda}_k} (\tilde{\lambda}) \right\vert \right) \leq \left\vert \frac{\gamma_2}{\gamma_1} \right\vert +  \bigg(  2 (n^* - 1) \vert \phi_1 \vert  + (2 n^* - 1) \vert \phi_2 \vert  \bigg) \tilde \eta.
\end{equation} 
Assume $|\gamma_2 / \gamma_1 | < 1$. By Lemma \ref{lemma:etabound}, we can always find  $\delta$ large enough to ensure that
\begin{equation*}
    \left\vert \frac{\gamma_2}{\gamma_1} \right\vert +  \bigg(  2 (n^* - 1) \vert \phi_1 \vert  + (2 n^* - 1) \vert \phi_2 \vert  \bigg) \tilde \eta < 1,
\end{equation*}
By Theorem 1 and Remark 1 in \cite{allen2024}, $g$ has a unique solution in $\tilde \Lambda^k$.
\end{proof}

\subsection{Proof of Theorem \ref{theo:theo2}}
\label{app:proofstheo2}
For $\alpha = 1/(\sigma - 1)$, the equilibrium system \eqref{eq:lambda_eq} becomes: 
\begin{equation*}
e^{\delta \sigmatil \sigma \lambda_i } = V^{\frac{1}{\beta}} \sum_{j = 1}^{n^*} T_{ij}^{1-\sigma} \bar{A}_i^{\tilde{\sigma}(\sigma-1)} \bar{A}_j^{\tilde{\sigma}\sigma} B_j(\lambda)^{-\frac{1}{\beta}} e^{ - \frac{\delta}{\beta} \sigmatil \gamma_2 \lambda_j} \quad i = 1, \dots n^*.
\end{equation*}
Also, in the proof of Proposition \ref{prop:lambdap} we derived the following expression for the potential weight at vacant business districts 
\begin{equation*}
    e^{\delta \sigmatil \sigma \lambda_p} =  V^{\frac{1}{\beta}} \sum_{j = 1}^{n^*} T_{pj}^{1-\sigma} \bar{A}_p^{\sigmatil(\sigma-1)}  \bar{A}_j^{\tilde{\sigma}\sigma}  B_j(\lambda)^{-\frac{1}{\beta}} e^{ - \frac{\delta}{\beta} \sigmatil \gamma_2}, \quad i = n^* + 1 , \dots , n.
\end{equation*}
We can summarize the spatial equilibrium with a single system of equations, 
\begin{equation*}
e^{\delta \sigmatil \sigma \lambda_i } = V^{\frac{1}{\beta}} \sum_{j = 1}^{n} T_{ij}^{1-\sigma} \bar{A}_i^{\tilde{\sigma}(\sigma-1)} \bar{A}_j^{\tilde{\sigma}\sigma} B_j(\lambda)^{-\frac{1}{\beta}} e^{ - \frac{\delta}{\beta} \sigmatil \gamma_2 \lambda_j} \quad i = 1, \dots n.
\end{equation*}
Let $\lambdatil_i$, $K_{ij}$ and $\tilde B_i$ be defined as in the proof of Theorem \ref{theo:theo1} for all $i, j = 1, \dots, n$. After taking logarithms, the equilibrium system in the transformed variables becomes:
\begin{equation}\label{eq:lambdatil_eq_n}
    \lambdatil_i = \log \sum_{j = 1}^{n} K_{ij} \tilde B_j(\lambdatil)^{-\frac{1}{\beta}} e^{ \frac{\gamma_2}{\gamma_1}
 \lambdatil_j } , \quad i, \dots, n.
\end{equation}
Consider the map $g: \R^n \to \R^n$, 
\begin{equation*}
    g_i(\lambdatil) = \log \sum_{j = 1}^{n} K_{ij} \tilde B_j(\lambdatil)^{-\frac{1}{\beta}} e^{ \frac{\gamma_2}{\gamma_1}
 \lambdatil_j } , \quad i = 1, \dots, n.
\end{equation*}
We study the fixed points of $g$ in $\R^n$.

\begin{proof}[Proof of Theorem \ref{theo:theo2}]
\textit{Part \ref{cond:2a}.} Fix one coordinate $y_{i_0} \in \Y$ and let $\hat \lambda_i$ and $\hat B_i$ be defined as in the proof of Theorem \ref{theo:theo1} for all $i = 1, \dots, n$. Consider the map $\hat g$,  
\begin{equation*}
    \hat g_i(\hat \lambda) = \log\frac{ \sum_{j = 1}^{n} K_{ij}  \hat B_j(\hat \lambda)^{-\frac{1}{\beta}} e^{ \frac{\gamma_2}{\gamma_1}
 \lambdatil_j } } { \sum_{j = 1}^{n} K_{i_0j} \hat B_j(\hat \lambda)^{-\frac{1}{\beta}} e^{ \frac{\gamma_2}{\gamma_1}
 \lambdatil_j } }.
\end{equation*}
Note that, for all $\hat \lambda \in \R^{n-1}$, 
\begin{equation*}
     \log \frac{ \min _jK_{ij}}{\max_j K_{i_0 j}} \leq \hat g_i(\hat \lambda) \leq \log \frac{ \max _jK_{ij}}{\min_j K_{i_0 j}}.
\end{equation*}
Therefore, $\hat g$ is a continuous function that maps a compact convex set into itself. By Schauder fixed point theorem, it has a fixed point in $\hat \Lambda^k$. Let $\hat \lambda^* \in \hat \Lambda^k$ be one such fixed point. Then $\tilde \lambda^*$ defined as $\tilde \lambda^*_i = \hat \lambda^*_i$ for $i\neq i_0$ and $\tilde \lambda^*_{i_0} =0$ is a fixed point of $g$. 

\textit{Proof of \ref{cond:2b}.}  The proof follows the same structure as the proofs of Lemmas \ref{lemma:gcontraction} and \ref{lemma:etabound} in the proof of Theorem \ref{theo:theo1}. That is, we show that $$\max_{i = 1, \dots, n}  \sup_{\lambdatil \in \tilde \Lambda^k} \sum_{k=1}^{n}\left|\frac{\partial g_i}{\partial  \lambdatil_k}(\lambdatil)\right|$$ is bounded and that, provided that $|\gamma_2 / \gamma_1| < 1$, this upper bound becomes lower than one as $\delta$ grows large. For all $i,j =1,\dots,n$, let 
\begin{equation}\notag
f_{ij}(\tilde{\lambda}) = K_{ij}\tilde{B}_j(\tilde{\lambda})^{-\frac{1}{\beta}}e^{\frac{\gamma_2}{\gamma_1}\tilde \lambda_j},
\end{equation}
so that $g_i(\tilde \lambda) = \log \sum_{j=1}^{n} f_{ij}(\tilde \lambda)$, and let
\begin{align*}
    S_{ik}(\lambdatil) &= \frac{\partial }{\partial \lambdatil_k} \tilde{B}_i(\tilde{\lambda})^{-\frac{1}{\beta}}  = \int_{\Gamma_{ik} } \left(\frac{\bar b(x)}{D(x, y_i)} \right)^{-\frac{1}{\beta}} \frac{\partial \omega_{ik}}{\partial \lambda_k} (x,\lambda) \cdot \nu_i(x) d\sigma(x).
\end{align*}
Let $\Y^*(\lambdatil) \subseteq \Y$ be the set of business districts with a positive commuting area,  $\Y^*(\lambdatil) = \{y_i \in \Y: |\Omega_i(\lambdatil)| > 0 \}$. Henceforth, we omit the dependency on $\lambdatil$. We have 
\begin{align*}
\sum_{k = 1}^{n} \left| \frac{\partial g_i}{\partial \tilde{\lambda}_k} \right| &= \sum_{k = 1}^{n} \left| \frac{ \sum_{j=1}^{n} \partial f_{ij} / \partial \tilde{\lambda}_k}{\sum_{j=1}^{n}  f_{ij}} \right|  \\
& = \sum_{y_k \in \Y^*} \left| \frac{ \sum_{y_j \in \Y^*} \partial f_{ij} / \partial \tilde{\lambda}_k}{\sum_{y_j \in \Y^*}  f_{ij}} \right| 
\leq \sum_{y_k \in \Y^*} \left| \frac{ \sum_{y_j \in \Y^*} \partial f_{ij} / \partial \tilde{\lambda}_k}{\sum_{y_j \in \Y^*}  f_{ij}} \right|,
\end{align*}
where in the second equality we have used the fact that $f_{ij} = 0$ whenever $|\Omega_j| = 0$, and $\partial f_{ij} / \partial \lambdatil_k$ whenever either $y_j$ or $y_k$ are not active. For all $k = 1, \dots, n$, we have
\begin{align*}
\frac{\partial f_{ik}}{\partial \tilde{\lambda}_k} &=
\left( \frac{\gamma_2}{\gamma_1}\tilde{B}_{k}(\lambdatil)^{-\frac{1}{\beta}} + S_{ik} \right) K_{ik} e^{\frac{\gamma_2}{\gamma_1}\lambdatil_k},  \\
\frac{\partial f_{ij}}{\partial \tilde{\lambda}_k} &= S_{jk} K_{ij} e^{\frac{\gamma_2}{\gamma_1}\lambdatil_j}
,  \quad  k \neq j.
\end{align*}
Using this derivatives into the previous inequality, we obtain
\begin{align*}
    \sum_{k = 1}^n \left|  \frac{\partial g_i}{\partial \lambdatil_k}  \right| &\leq \left| \frac{\gamma_2}{\gamma_1} \right| +  \sum_{y_k \in \Y^*}  \sum_{y_{\ell} \in \Y^*} \frac{ K_{i \ell} e^{\frac{\gamma_2}{\gamma_1} \lambdatil_{\ell}} }{\sum_{y_j \in \Y^*} f_{ij}} |S_{\ell k}|
    \end{align*}
Next, we use some basic properties of the shape derivative (see Appendix \ref{app:shapeder}). In particular, $S_{ii} = \sum_{y_k \in \mathcal{N}_i} |S_{ik}|$, as the own-weight effect is the sum of the cross-weight effects acting, with a negative sign, on the opposite side of each border segment. Because $|S_{ik}| \leq S$ for $i \neq k$ and the number of neighbors is at most $n-1$, we find
\begin{align*}
    \sum_{k = 1}^n \left|  \frac{\partial g_i}{\partial \lambdatil_k}  \right| &\leq  \left| \frac{\gamma_2}{\gamma_1} \right| +  (2n - 1) S \frac{\sum_{y_k \in \Y^*} K_{ik} e^{\frac{\gamma_2}{\gamma_1} \lambdatil_k} }{\sum_{y_j \in \Y^*} K_{ij}\tilde{B}_j(\tilde{\lambda})^{-\frac{1}{\beta}}e^{\frac{\gamma_2}{\gamma_1}\tilde \lambda_j}}.
\end{align*}
Because $|\Omega_i| > 0$ for at least one $y_i \in \Y$, for all $\Y^* \subseteq \Y$ we can choose $\epsilon > 0$ and $y_i \in \Y^*$ such that 
\begin{equation*}
    \sum_{k = 1}^n \left|  \frac{\partial g_i}{\partial \lambdatil_k}  \right| \leq \left| \frac{\gamma_2}{\gamma_1} \right| +  (2n - 1) \frac{S}{\int_{B_{\epsilon}(y_i)} \left(\frac{\bar b(x)}{D(x, y_i)} \right)^{-\frac{1}{\beta}} \d x },
\end{equation*}
where $B_{\epsilon}(y_{i_0})$ is an open ball of radius $\epsilon$ and centered in $y_i$. By Lemmas \ref{lemma:Bbound} and \ref{lemma:etabound}, we have that
\begin{equation*}
\frac{S}{\int_{B_{\epsilon}(y_i)} \left(\frac{\bar b(x)}{D(x, y_i)} \right)^{-\frac{1}{\beta}} \d x } \leq 
 \frac{ K }{c^{\prime \prime}} \left( \frac{\bar b_{max}}{\bar b_{min}} \right)^{-\frac{1}{\beta}}\delta^2  e^{\frac{\delta}{\beta} \epsilon},
\end{equation*}
where the constants $K$ and $c^{\prime \prime}$ are given in the proofs of Lemma \ref{lemma:etabound}. Since $\beta < 0$, the right-hand side approaches zero as $\delta \to +\infty$. Assume $|\gamma_2 / \gamma_1 | < 1$. Then, we can always find  $\delta$ large enough to ensure that
\begin{equation*}
    \max_{i = 1, \dots, n}  \sup_{\lambdatil \in \tilde \Lambda^k} \sum_{k=1}^{n}\left|\frac{\partial g_i}{\partial  \lambdatil_k}(\lambdatil)\right| < 1.
\end{equation*}
By Theorem 1 and Remark 1 in \cite{allen2024}, $g$ has a unique solution in $\tilde \Lambda^k$.
\end{proof}

\subsection{Proof of Lemma \ref{lemma:multipleY}}\label{app:proofsmultipleY}

\begin{proof}[Proof of Lemma \ref{lemma:multipleY}]
Under Assumptions \ref{ass:Ab}-\ref{ass:d}, and with $\gamma_1 \neq 0$, the results of Theorem \ref{theo:theo1} apply. Thus we verify that the sufficient condition \eqref{eq:suffcondexist} for the existence of a spatial equilibrium in $\Y^{**} \subset \Y$ is satisfied under the assumptions of the lemma. To simplify the exposition, we only present the proof for the case $\gamma_1 > 0$. Also, let $C = c + \sigmatil | \phi_1| \log \left( \bar b_{max}/\bar b_{min} \right)$.  Under Assumption \ref{ass:T2}, we can use Remark \ref{remark:suffcondexist} to write the sufficient condition as
\begin{align*}
         C + \sigmatil (\sigma - 1) \left| \log  \frac{\bar A_i}{\bar A_j} \right|  < \left( -\frac{\delta}{\beta} \sigmatil \gamma_1 - \tau(\sigma - 1) \right) d_i(y_j),
\end{align*}
for all $y_i, y_j \in \Y^{**}$ with $y_i \neq y_j$. By hypothesis, this sufficient condition holds for the $n^*$ business districts in $\Y^*$. Therefore, the inequalities are satisfied for all $y_i, y_j \in \Y^* \bigcap \Y^{**}$ with $y_i \neq y_j$, and we only need to verify the ones where $y_p$ appears:
\begin{align}\label{eq:gammaboundsyp1}
         C + \sigmatil (\sigma - 1) \left| \log \frac{\bar A_p}{\bar A_j} \right|  < \left( -\frac{\delta}{\beta} \sigmatil \gamma_1 - \tau(\sigma - 1) \right) d_p(y_j),
\end{align}
for all $y_j \in \Y^{**}$, with $y_j \neq y_p$. By the triangle inequality (Assumption \ref{ass:trineq}), $d_c(y_j) - d_c(y_p) \leq d_p(y_j)$. Therefore, it is  sufficient to show that
\begin{align*}
    C + \sigmatil (\sigma - 1) \left| \log \frac{\bar A_p}{\bar A_j} \right| &< \left( -\frac{\delta}{\beta} \sigmatil \gamma_1 - \tau(\sigma - 1) \right)\left(d_c(y_j) - d_c(y_p)  
    \right).
\end{align*}
Again, by hypothesis
\begin{align*}
    C + \sigmatil (\sigma - 1) \left| \log \frac{\bar{A}_{c}}{\bar{A}_{j}}  \right| & < \left( -\frac{\delta}{\beta} \sigmatil \gamma_1 - \tau(\sigma - 1) \right) d_c(y_j).
\end{align*}
Then, for $ \bar A_c / \bar A_p $ close to $1$ and $d_c(y_p) (=d_p(y_c))$ small, \eqref{eq:gammaboundsyp1} will be satisfied too.
\end{proof}

\subsection{Proof of Theorem \ref{theo:theo3}}
\label{app:proofstheo3}

\begin{lemma}\label{lemma:Bbound2}
Suppose that Assumptions \ref{ass:Ab}-\ref{ass:dmetric} hold, and let $k \in (0,1)$. Then, for all $\lambda \in \Lambda^k$ and $y_i, y_j \in \Y^*$, $y_i \neq y_j$, $| \log B_i(\lambda) / B_j(\lambda) |$ has an upper bound that is decreasing in $d_{min}$.
\end{lemma}
\begin{proof}
Take $\underline{\lambda}^i \in \Lambda^k$ such that $\lambda_i - \lambda_j = -k d_j(y_i)$ for all $y_j \in \Y^*$ with $y_j \neq y_i$. Note that $\Omega_i(\underline{\lambda}^i) \subseteq \Omega_i(\lambda)$ for all $\lambda \in \Lambda^k$. Along the bisector $B_{ij}(\underline{\lambda}^i)$ between $y_i$ and $y_j$, we have: 
\begin{equation*}
    d_i(x) - d_j(x) = - k d_j(y_i), \quad \text{all } x \in \Gamma_{ij}(\underline{\lambda}^i).
\end{equation*}
By Assumption \ref{ass:dmetric}, the triangle inequality implies: 
\begin{equation} \label{eq:triang2}
    d(y_i,x) \geq \frac{(1- k)}{2} d(y_i,y_j) \quad \text{all } x \in \Gamma_{ij}(\underline{\lambda}^i).
\end{equation}
Let $B
_\varepsilon(y_i)$ denote an open ball centered in $y_i$ of radius $\varepsilon$ in the Euclidean distance. Also, let $c, C$, with $0 < c < C$ be the constants from Assumption \ref{ass:doublelip}. Using Assumption \ref{ass:doublelip} and inequality \eqref{eq:triang2}, we have
\begin{equation*}
||x - y_i || \geq\frac{d(x, y_i)}{C} \geq \frac{(1- k)}{2C} d(y_i,y_j) \quad \text{all } x \in \Gamma_{ij}(\underline{\lambda}^i).
\end{equation*}
Thus, we can set $\epsilon = \frac{1-k}{2C} d_{min} $ such that $B_\varepsilon (y_i) \subseteq \Omega_i(\lambda)$ for all $\lambda \in \Lambda^k$ and all $y_i \in \Y^*$. To simplify the exposition, let use define: 
\begin{align*}
    \mathcal{I}(\epsilon, a) &=  2\pi \left[ \left( 1 - e^{\frac{\delta a\varepsilon}{\beta}} \right) \frac{ \beta^2}{(\delta a)^2}+\varepsilon  e^{\frac{\delta a \varepsilon}{\beta}} \frac{\beta}{\delta a} \right], \quad a \in \{c,C\}, \\
    \mathcal{J}(\epsilon) &= \frac{\max_{y_i \in \Y^*}|\Omega_i(\lambda) \setminus B_{\epsilon}(y_i)|}{e^{-\frac{\delta}{\beta}c\epsilon}}.
\end{align*}
Using the calculations in Lemma \ref{lemma:Bbound}, we find
\begin{align*}
\bar b_{min} \mathcal{I}(\epsilon, C)^{-\beta} \leq B_i(\lambda)  \leq \bar b_{max} \left( \mathcal{I}(\epsilon, c) + \mathcal{J}(\epsilon) \right)^{-\beta}.
\end{align*}
It follows that
\begin{equation*}
     \frac{B_i(\lambda)}{B_j(\lambda)} \leq \frac{\bar b_{max}}{\bar b_{min}} \left(\frac{\mathcal{I}(\epsilon, c) + \mathcal{J}(\epsilon) }{\mathcal{I}(\epsilon, C)} \right)^{-\beta},
\end{equation*}
for all $y_i, y_j \in \Y^*$, $y_i \neq y_j$. Finally, we have that 
\begin{align*}
    \mathcal{I}^{\prime}(\epsilon, a) &= 2 \pi \epsilon e^{\frac{\delta a \epsilon}{\beta}} > 0, \\
    \mathcal{I}^{\prime}(\epsilon, c) + \mathcal{J}^{\prime}(\epsilon) &= \frac{\delta}{\beta}c e^{\frac{\delta}{\beta}c \epsilon} \left(\max_{y_i \in \Y^*} |\Omega_i(\lambda)| - \pi \epsilon^2 \right) <0,
\end{align*}
noting that $|B_{\epsilon}(y_i) |= \pi \epsilon^2$ for all $i=1, \dots, n^*$.
\end{proof}

\begin{lemma}\label{lemma:etabound2}
Suppose that Assumptions \ref{ass:Ab}-\ref{ass:dmetric} hold. Then, $\eta$ is decreasing in $d_{min}$.
\end{lemma}
\begin{proof}
Recall that 
\begin{equation*}
    \eta_{ik} = - \beta \frac{\int_{\Gamma_{ik}} \left(\frac{\bar b(x)}{D(x,y_i)}\right)^{-\frac{1}{\beta}} \frac{\partial \omega_{ik}}{\partial \lambda_k }(x, \lambda) \cdot \nu(x) \d \sigma(x) }{\int_{\Omega_i} \left(\frac{\bar b(x)}{D(x,y_i)}\right)^{-\frac{1}{\beta}} \d x}.
\end{equation*}

Let $\epsilon = \frac{1-k}{2} d_{min}$. In Lemma \ref{lemma:etabound}, we showed that \begin{align*}
\left|\int_{\Gamma_{ik}} \left(\frac{\bar b(x)}{D(x,y_i)}\right)^{-\frac{1}{\beta}} \frac{\partial \omega_{ik}}{\partial \lambda_k }(x, \lambda) \cdot \nu(x) \d \sigma(x)\right|  &\leq \int_{\Gamma_{ik}}  \left(\frac{\bar b(x)}{D(x,y_i)}\right)^{-\frac{1}{\beta}} \left| \frac{\partial \omega_{ik}}{\partial \lambda_k }(x, \lambda) \cdot \nu(x) \right| \d \sigma(x) \notag \\ 
& \leq \bar b_{max}^{-\frac{1}{\beta}} K e^{\frac{\delta}{\beta} \epsilon },
\end{align*}
where $K$ is a constant defined in Lemma \ref{lemma:etabound}. In
Lemma \ref{lemma:Bbound2}, we showed that
\begin{equation*}
    \tilde B_i(\lambdatil)^{-\frac{1}{\beta}} = \int_{\Omega_i} \left(\frac{\bar b(x)}{D(x,y_i)}\right)^{-\frac{1}{\beta}} \d x \geq 2\pi \left[ \left( 1 - e^{\frac{\delta C\varepsilon}{\beta}} \right) \frac{ \beta^2}{(\delta C)^2}+\varepsilon  e^{\frac{\delta C \varepsilon}{\beta}} \frac{\beta}{\delta C} \right],
\end{equation*}
where $C$ is a constant from Assumption \ref{ass:doublelip}. By combining these bounds, we get: 
\begin{equation*}
    |\eta_{ik}| \leq \frac{\bar b_{max}^{-\frac{1}{\beta}} K e^{\frac{\delta}{\beta} \epsilon } }{2\pi \left[ \left( 1 - e^{\frac{\delta C\varepsilon}{\beta}} \right) \frac{ \beta^2}{(\delta C)^2}+\varepsilon  e^{\frac{\delta C \varepsilon}{\beta}} \frac{\beta}{\delta C} \right] }.
\end{equation*}
Since $\beta < 0$, in the right-hand side of this expression the numerator decreases with $\epsilon$, while the denominator increases with $\epsilon$. Because the right-hand side is independent both of $i$ and $\lambdatil$, the same inequality holds for the maximum with respect to $i$ in $\{1, \dots, n^*\}$ and $\lambdatil$ in the closed set $\tilde \Lambda^k$.
\end{proof}

\begin{proof}[Proof of Theorem \ref{theo:theo3}]
We only provide a sketch of the proof, as it follows the same logic of the proof of Theorem \ref{theo:theo1}. To establish existence, we use the inequalities \eqref{eq:gammabounds} in Lemma \ref{lemma:gmap}. Fix one coordinate $y_{i_0} \in \Y^*$, and define the changes variables $\lambdatil$ and $\hat \lambda$, as well as the sets $\tilde \Lambda^k$ and $\hat \Lambda^k$, and the maps $g$ and  $\hat g$, as in the proof of Theorem \ref{theo:theo1}. Using the a priori bounds on $\{\bar{A}_i\}$ from Assumption \ref{ass:Ab}, the bounds on $\{B_i\}$ from Lemma \ref{lemma:Bbound2}, and the bounds on the market access term \eqref{eq:suffcondexistT2} in Remark \ref{remark:suffcondexist}, it is clear that, as long as $ - \frac{\delta}{\beta} \sigmatil |\gamma_1| -  \tau (\sigma - 1) > 0$, there exists $d^{*}>0$ such that for $d_{min} > d^{*} $ all the inequalities \ref{eq:gammabounds} in Lemma \ref{lemma:gmap} will be satisfied. Then, $\hat g$ is a continuous function that maps a nonempty convex compact subset of $\R^{n^*-1}$ and, by Schauder fixed point theorem, it has a fixed point. To prove uniqueness, we use Lemma \ref{lemma:gcontraction} and \ref{lemma:etabound2}. Together, these lemmas imply that, as long as $|\gamma_2 / \gamma_1| < 1$, there exists $d^{**} > d^{*}$ such that, for $d_{min} > d^{**}$ the map $g$ is a contraction that maps a complete metric space onto itself. By Banach fixed point theorem, it has a unique fixed point. 
\end{proof}

\subsection{Proof of Lemma \ref{lemma:gmap2}}\label{app:proofsgmap2}

\begin{proof}
Fix a coordinate $y_{i_0} \in \Y^*$ such that  $\max_{i \neq i_0} d_i(y_{i_0}) = r$. Let $\hat \lambda$, $\hat g$, and $\hat \Lambda^k$ be defined as in the proof of Theorem \ref{theo:theo1}. Because Assumptions \ref{ass:T} and \ref{ass:T2} imply that $d_i(y_j) = d_j(y_i)$ for all $y_i, y_j \in \Y$, we can write 
\begin{equation*}
\begin{split}
\hat \Lambda^k = \bigg\{ \hat \lambda \in \R^{n^* -1}: & \\
|\hat \lambda_i | & \leq - \frac{\delta}{\beta} \sigmatil | \gamma_1 | k d_{i}(y_{i_0}), \text{ all } i = 1,\dots, n^*, i \neq i_0 ,
\\
| \hat \lambda_i &- \hat \lambda_j | \leq - \frac{\delta}{\beta} \sigmatil | \gamma_1 | k d_i(y_j),  \text{ all } i,j = 1,\dots, n^*, i,j \neq i_0,  i \neq j \bigg\}.
\end{split}
\end{equation*}
Choose $\hat \lambda \in \hat \Lambda^k$. We show that under the assumptions of the lemma, $\hat g: \hat \Lambda^k \to \hat \Lambda^k$, that is
\begin{align*}
    | \hat g_i(\hat \lambda) | & \leq - \frac{\delta}{\beta} \sigmatil | \gamma_1 | k d_{i}(y_{i_0}), \text{ all } i = 1,\dots, n^*, i \neq i_0 ,
\\
| \hat  g_i(\hat \lambda) - \hat  g_j(\hat \lambda) | &\leq - \frac{\delta}{\beta} \sigmatil | \gamma_1 | k d_i(y_j),  \text{ all } i,j = 1,\dots, n^*, i,j \neq i_0,  i \neq j.
\end{align*}
 A sufficient condition for these inequalities to hold is
\begin{equation*}
 \sigmatil(\sigma - 1) \left| \log\frac{\bar A_i}{\bar A_j} \right| + \sigmatil |\phi_1| \left| \log \frac{\hat B_i(\hat \lambda)}{\hat B_j(\hat \lambda)} \right| + \left| \log \frac{\sum_{k = 1}^{n^*} T_{ik}^{1-\sigma} \bar{A}_k^{\tilde{\sigma}\sigma} \hat B_k(\hat \lambda)^{-\frac{1}{\beta}} e^{ - \frac{\delta}{\beta}
    \sigmatil \gamma_2  \hat \lambda_k}}{\sum_{k = 1}^{n^*} T_{jk}^{1-\sigma} \bar{A}_k^{\tilde{\sigma}\sigma} \hat B_k(\hat \lambda)^{-\frac{1}{\beta}} e^{ - \frac{\delta}{\beta}
    \sigmatil \gamma_2 \hat \lambda_k}} \right| \leq - \frac{\delta}{\beta} \sigmatil | \gamma_1 | k d_i(y_j),  
\end{equation*}
for all $i,j = 1,\dots, n^*$, and $ i \neq j$. Due to Assumption \ref{ass:Ab}, $\bar A_i / \bar A_j$ is bounded above and below. Furthermore, due to Assumptions \ref{ass:T} and \ref{ass:T2}, and using Remark \ref{remark:suffcondexist}, 
\begin{equation*}
    \log \frac{\sum_{k = 1}^{n^*} T_{ik}^{1-\sigma} \bar{A}_k^{\tilde{\sigma}\sigma} \hat B_k(\hat \lambda)^{-\frac{1}{\beta}} e^{ - \frac{\delta}{\beta}
    \sigmatil \gamma_2  \hat \lambda_k}}{\sum_{k = 1}^{n^*} T_{jk}^{1-\sigma} \bar{A}_k^{\tilde{\sigma}\sigma} \hat B_k(\hat \lambda)^{-\frac{1}{\beta}} e^{ - \frac{\delta}{\beta}
    \sigmatil \gamma_2 \hat \lambda_k}} \leq \tau (\sigma -1) d_i(y_j).
\end{equation*}
Next, we focus on the term $ \hat B_i(\hat \lambda) / \hat B_j (\hat \lambda)$. For all $i = 1, \dots, n^*$, and all $k \neq i_0$, let 
\begin{equation*}
\hat \eta_{ik}(\hat \lambda) = \frac{\partial \hat B_i(\hat \lambda)}{\partial \hat \lambda_k} \frac{1}{\hat B_i(\hat \lambda)}.
\end{equation*}
According to the mean value theorem, for any two vectors $\hat{\lambda}$ and $\hat{\lambda}^{\prime}$ in $\R^{n*-1}$ there exists $t \in [0,1]$ such that $\hat{\lambda}^t = t\hat \lambda + (1-t) \hat \lambda^{\prime}$ solves:  
\begin{align*}
    (\log \hat B_i)(\hat \lambda) - (\log \hat B_i)(0) &= \sum_{k \neq i_0} \frac{\partial (\log \hat B_i)(\hat \lambda^t)}{\partial \hat \lambda_k} \hat \lambda_k \iff \\
     | (\log \hat B_i)(\lambdatil) - (\log \hat B_i)(0) | & \leq \max_{i = 1,\dots,n^*} \max_{\hat \lambda^t \in \hat \Lambda^k} \left( \sum_{k \neq i_0} \left| \hat \eta_{ik}(\hat \lambda^t) \right| \right) \| \hat \lambda \|_{\infty} \\
     & \leq (n^* - 1) \hat \eta \| \hat \lambda \|_{\infty}.
\end{align*}
It follows that
\begin{align*}
 \log \hat B_i(0) - (n^* - 1) \hat \eta \| \hat \lambda \|_{\infty} \leq \log \hat B_i(\hat \lambda) \leq  \log \hat B_i(0) + (n^* - 1) \hat \eta \| \hat \lambda \|_{\infty},
\end{align*}
which implies 
\begin{align*}
    \left| \log \frac{\hat B_i(\hat \lambda)}{\hat B_j(\hat \lambda)} \right| &\leq \left|\log \frac{\hat B_i(0)}{\hat B_j(0)}\right| + 2 (n^* - 1) \hat \eta  \| \hat \lambda \|_{\infty} \\
     &\leq \left|\log \frac{\hat B_i(0)}{\hat B_j(0)}\right| - 2 (n^* - 1) \hat \eta \frac{\delta}{\beta} \sigmatil |\gamma_1| r,
\end{align*}
where the second inequality uses the fact that $\hat \lambda \in \hat \Lambda^k$ and the choice of $y_{i_0}$. By combining these upper bounds, we find that a sufficient condition for $\hat g: \hat \Lambda^k \to \hat \Lambda^k$ is: 
\begin{multline*}
    \sigmatil(\sigma - 1) \left| \log  \frac{\bar A_i}{\bar A_j} \right| + \sigmatil |\phi_1|\left| \log  \frac{\hat B_i(0)}{\hat B_j(0)} \right| - 2 (n^* - 1) \hat \eta \frac{\delta}{\beta} \sigmatil |\gamma_1| r \\ \leq \left( - \frac{\delta}{\beta} \sigmatil |\gamma_1| -  \tau (\sigma - 1) \right) d_i(y_j),
\end{multline*}
for all $y_i, y_j \in \Y^*$ with $y_i \neq y_j$. In the original $\lambda$ variables, these inequalities become
\begin{equation*}
    \sigmatil(\sigma - 1) \left| \log \frac{\bar A_i}{\bar A_{j}} \right| + \sigmatil |\phi_1| \left|\log  \frac{B_i(0)}{B_{j}(0)} \right| - 2 \beta \eta r \leq \left( - \frac{\delta}{\beta} \sigmatil |\gamma_1| -  \tau (\sigma - 1) \right) d_i(y_{j}),
\end{equation*}
for all $y_i, y_j \in \Y^*$, with $y_i \neq y_j$. Therefore, if \eqref{eq:gmap2} holds, $\hat{g}$ is a continuous function that maps a nonempty convex compact subset of $\R^{n^*-1}$ onto itself. By Schauder fixed point theorem, it has a fixed point in $\hat \Lambda^k$. 
\end{proof}

\section{A model with home consumption}\label{app:homeconsumption}

In the model presented in Section \ref{sec:model}, we assumed that agents consume the bundle of differentiated varieties at their commuting destination. Suppose, instead, that consumption takes place at their residential location. Let us define the trade cost function: $T: \X \times \X \to [1, \infty)$, such that, such that each unit of the good shipped from $x$ results in $1/T(x, x^{\prime})$ units reaching their destination at $x^{\prime}$. The unit price of a variety produced at $y_i$ and purchased at $x$ is $p^i(x) = T(y_i, x) p^j$, and the price index at $x$ is: 
\begin{equation*}
    P(x) = \left( \sum_{j = 1}^{n^*} T(y_j, x)^{1-\sigma} (p^j)^{1-\sigma} \right)^{\frac{1}{1-\sigma}}.
\end{equation*}
To make the model tractable, we assume that the residents of $x \in \Omega_i$ must source the imported varieties from their respective business district $y_i \in \Y$, giving us $T(y_i, x)  = T(y_i, y_j) T(y_j, x)$. For a given commuting pattern, we can express the total demand for the variety produced at $y_i$ from commuting area $\Omega_j$ as follows:
\begin{equation*}
     \int_{\Omega_j} T(y_i,x) (p^i(x))^{-\sigma} P(x)^{\sigma-1} w_j \ell(x) =     T(y_i,y_j) (p^i)^{-\sigma} P(x)^{\sigma-1} w_j  \int_{\Omega_j} \ell(x).
\end{equation*}
As a result, the market clearing conditions for the manufacturing varieties become identical to \eqref{eq:gmc} with $C^i_j = T(y_i,y_j) (p^i)^{-\sigma} P(x)^{\sigma-1} w_j$ for all $y_i, y_j \in \Y^*$. 

At the optimal consumption choice, we can express the welfare of an agent residing at $x \in \X$ and commuting to $y_i \in \Y^*$ with the indirect utility function:
\begin{equation*}
	V(x, y_i) =  \frac{b(x) w_i}{T(y_i, x) D(x, y_i) P_i}.
\end{equation*}
Agents solve problem \eqref{eq:maxV} with this indirect utility. To make progress, we define the continuous distance function $d: \X \times \X \to \R_{+}$ and impose the equivalents of Assumptions \ref{ass:D} and \ref{ass:T2} on commuting and trade costs, that is, $T(x, x^{\prime}) = e^{\tau d(x, x^{\prime})}$ and $D(x, x^{\prime}) = e^{\delta d(x, x^{\prime})}$ for $\tau,\delta > 0$ and all $x, x^{\prime} \in \X$. Because we will maintain the assumption that trade costs are symmetric, we are implicitly assuming that the distance function $d$ is symmetric too. 

For a fixed residential location $x \in \X$, the choice over $\Y^*$ gives an additively weighted Voronoi tessellation with weights: 
\begin{equation*}
    \lambda_i= \frac{1}{\delta +\tau} \log \frac{w_i}{P_i}, \quad i = 1,\dots, n^*.
\end{equation*}
For a fixed tessellation, the choice of a residential location $x \in\ X$ gives: 
\begin{align*}
\ell(x) &= \frac{\left( \bar{b}(x)/(T(y_i,x) D(x, y_i))\right)^{-\frac{1}{\beta}}}{\int_{\Omega_i}\left( \bar{b}(x)/(T(y_i,x) D(x, y_i))\right)^{-\frac{1}{\beta}}\d x } \int_{\Omega_i} \ell(x)\d x, \quad x \in \Omega_i , y_i \in \Y^*,
\end{align*}
Therefore, equation \eqref{eq:L} is unchanged, except that $B_i$ is now defined as
\begin{equation*}
    B_i = \left[ \int_{\Omega_i}\left( \frac{ \bar{b}(x)}{T(y_i,x) D(x, y_i)}\right)^{-\frac{1}{\beta}}\d x \right]^{-\beta} \; i = 1, \dots, n.
\end{equation*}
Finally, equation \eqref{eq:lambda_eq} becomes: 
\begin{equation*}
	e^{-\frac{\delta + \tau}{\beta} \tilde{\sigma}\gamma_1 \lambda_i } =  \sum_{j = 1}^{n^*} T_{ij}^{(1-\sigma)} \bar{A}_i^{\tilde{\sigma}(\sigma-1)} \bar{A}_j^{\tilde{\sigma}\sigma} B_i(\lambda)^{\sigmatil \phi_1} B_j(\lambda)^{\sigmatil \phi_2} e^{ -\frac{\delta  + \tau}{\beta} \tilde{\sigma}\gamma_2 \lambda_j },
\end{equation*}
with $\phi_1$ and $\phi_2$ as in the main text. By studying the solutions of this equation, we can characterize the existence and uniqueness properties of an equilibrium with $n^*$ cities.  

\section{A two-sector model}\label{app:twosector}

We augment the model presented in Section \ref{sec:model} with an agricultural sector.  Each location $x \in \X$ is endowed with the technology to produce an homogeneous agricultural good. We refer to the differentiated sector as the manufacturing sector. Agents employed in agriculture, i.e., farmers, reside at the production location and are entitled to an equal share of the local output. A farmer in $x\in \X$ consumes a fraction of her production locally and ships the surplus to an optimally chosen business district $y_i \in \Y$ subject to shipping cost $D(x, y_i)$. At the ``commuting'' destination $y_i$, farmers also purchase and consume the bundle of differentiated manufacturing varieties. The agents employed in manufacturing reside, work, and consume in the business districts. 

Agents order consumption baskets according to the utility function 
\begin{equation*}
    U = \frac{1}{\kappa_\mu} \left(\sum_{j = 1}^{n^*} (c^j)^{\frac{\sigma - 1}{\sigma}} \right)^{\frac{\sigma \mu}{\sigma-1} }(c^a)^{1- \mu}, \quad \text{ with } 0 < \mu < 1,
\end{equation*}
where $c^a$ denotes the consumed quantity of the agricultural good and $\kappa_\mu = \mu^{\mu} (1-\mu)^{1-\mu}$ is a constant that simplifies the algebra. Agricultural output per capita is given by 
\begin{equation*}
    y^A(x) = \bar b(x) l(x)^{\tilde{\beta}}, \quad x \in \X, \quad \text{ with } \tilde{\beta} < 0.  
\end{equation*}
where $\bar{b}(x)$ and $\ell(x)$ denote, respectively, the agricultural productivity and the density of farmers at $x$. Thus, there are decreasing returns to labor in the agricultural sector. 

Using two-stage budgeting to solve the consumer's problem, we obtain the following demand functions for a farmer living at $x$ and commuting to $y_i$:
\begin{align*}
    C^a(x) &= (1 - \mu) y^a(x), \\
    C^j_i(x) &= \mu (p^j_i)^{-\sigma} P_i^{\sigma - 1} \frac{ p^a_i y^a(x)}{D(x, y_i)} , \quad j = 1, \dots, n^* ,
    \end{align*}
where $p^a_i$ denotes the unit price of the agricultural good at $y_i$, and $P_i$ is given by equation $\eqref{eq:P}$. For a manufacturing worker at $y_i$, the demand functions are:
    \begin{align*}        
    C^a_i &= (1-\mu)\frac{ w_i }{p^a_i}, \\
    C^j_i &= \mu (p^j_i)^{-\sigma} P_i^{\sigma - 1} w_i , \quad j = 1, \dots, n^*.
\end{align*}
Agricultural and manufacturing markets clear when
\begin{align*}
& \int_{\Omega_i} \frac{ (y^a(x) - C^a(x) ) }{D(x,y_i)} \ell(x) \d x = C^a_i L_i \\
&Y_i = \sum_{j = 1}^{n^*} \left( C^i_j L_j + \int_{\Omega_j}C^i_j(x)\ell(x) \d x \right), \quad i = 1,\dots,n.
\end{align*}
For known geography and parameters, a given commuting pattern $\{\Omega_i\}_{y_i \in \Y^*}$ and $\ell$, and a given sectoral allocation $\{ \int_{\Omega_i} \ell(x) \d x, L_i \}$, we can describe the market equilibrium with a system of $2 \times n^*$ equations: the $n^*$ equations in \eqref{eq:marketeq}, and
\begin{equation}\label{eq:amc}
    \mu \int_{\Omega_i} \frac{y^a(x)}{D(x, y_i)} \ell(x) \d x = (1 - \mu) \frac{ w_i L_i }{p^a_i}, \quad \text{ for } i = 1, \dots, n^*.
\end{equation}
System of equations \eqref{eq:marketeq} takes the same form as in the one-sector model because the expenditures of agricultural and manufacturing workers are proportional when preferences are Cobb-Douglas and agricultural markets clear. As a result, we can use \eqref{eq:marketeq} to solve for wages and then use \ref{eq:amc} to compute the agricultural price at all business districts.  

At the optimal consumption choice, we can express the welfare of farmers and workers, respectively, with the indirect utility functions
\begin{equation*}
	V(x, y_i) =  \frac{p^a_i y^a(x)}{D(x, y_i)^{\mu} P_i^{\mu} (p^a_i)^{1-\mu}}, \quad V_i =  \frac{w_i}{P_i^{\mu} (p^a_i)^{1-\mu}},
\end{equation*}
for all $x \in \X$ and $y_i \in Y^*$. 

Agents make optimal location choices to maximize their welfare. Farmers choose a residential location $x \in \X$ and a commute destination $y_i \in \Y$, while manufacturing workers choose a business district in $y_i \in \Y$. All agents choose the employment sector that provides the highest level of welfare. We can write these choices as:
\begin{equation*}
\max \left\{ \max_{x \in \X, y_i \in \Y^*} V(x, y_i), \max_{y_i \in \Y} V_i \right\}.
\end{equation*} 
Because all business districts in $\Y^*$ are active by construction and $\tilde{\beta} < 0$, all locations and sectors attract a positive measure of agents and therefore offer the same level of welfare at the optimum (otherwise, agents would relocate to another location or sector). Thus, welfare equalization between employment sectors implies
\begin{equation*}
\frac{p^a_i \bar b(x) \ell(x)^{\tilde{\beta}}}{D(x,y_i)^{\mu}} = w_i \quad \text{ for all } x \in \Omega_i \text{ and } y_i \in \Y^*,
\end{equation*}
Next, consider the farmer's location problem. For a given $\ell$, the choice over $\Y^*$ delivers an additively weighted Voronoi tessellation with weights: 
\begin{equation*}
    \lambda_i = \frac{1}{\delta} \log \frac{p^a_i}{P_i}, \quad i = 1, \dots, n^*.
\end{equation*}
Fix $i\in \{1,\dots,n^*\}$. For a given tessellation, the choice over $\Omega_i \subseteq \X$ gives 
\begin{equation*}
    \frac{\ell(x)}{D(x,y_i)^{1-\mu}} = \frac{ (\bar b(x) / D(x,y_i)^{\mu - \tilde{\beta}(1-\mu)} )^{-\frac{1}{\tilde{\beta}}} }{\int_{\Omega_i} \big(\bar b(x) / D(x,y_i)^{\mu - \tilde{\beta}(1-\mu)} \big)^{-\frac{1}{\tilde{\beta}}} \d x} \int_{\Omega_i(\lambda)} \frac{\ell(x)}{D(x,y_i)^{1-\mu}}. 
\end{equation*}

By combining welfare equalization between sectors and across rural locations with the market clearing condition for the agricultural good, we  obtain: 
\begin{align*}
    \mu \int_{\Omega_i} \frac{\ell(x)}{D(x,y_i)^{1-\mu}} &= (1-\mu) L_i, \quad \text{ and} \\
    p^A_i &= \left( \frac{1-\mu}{\mu} \right)^{-\tilde{\beta}} \frac{L_i^{-\tilde{\beta}}}{B_i(\lambda)} w_i,
\end{align*}
where we have defined: 
\begin{equation*}
    B_i(\lambda) = \left[\int_{\Omega_i} \left( \frac{\bar b(x) }{ D(x,y_i)^{\mu - \tilde{\beta}(1-\mu)} } \right)^{-\frac{1}{\tilde{\beta}}} \d x \right]^{-\tilde{\beta}}.
\end{equation*}
We can interpret $B_i$ as the aggregate agricultural productivity of rural hinterland $\Omega_i$, after welfare has equalized among farmers and manufacturing workers.
Using these results, we can write the indirect utility of manufacturing workers as
\begin{equation*}
    V_i =  \left(\frac{1-\mu}{\mu} \right)^{ \left(\frac{1-\mu}{\mu} \right)\tilde \beta}B_i(\lambda)^{1-\mu}\left(\frac{w_i}{P_i}\right)^{\mu} L_i^{\tilde{\beta}(1-\mu)},
\end{equation*}
 Finally, the location choice of manufacturing workers implies $V_i = V$ for all $y_i \in \Y^*$, where the scalar $V>0$ denotes the common level of welfare across business districts (and the entire economy). By combining this condition with the expression for the price index and the market clearing condition for manufacturing varieties, we obtain: 
	\begin{align*}
	\label{eq:L} L_i^{\tilde{\sigma}\gamma_1} &= 
    \left(\frac{1-\mu}{\mu} \right)^{(\sigma - 1)  \left(\frac{1-\mu}{\mu}  \right) \tilde \beta} V^{\frac{1-\sigma}{\mu}} \sum_{j = 1}^{n^*} T_{ij}^{1-\sigma} \bar{A}_i^{\tilde{\sigma}(\sigma-1)} \bar{A}_j^{\tilde{\sigma}\sigma} B_i(\lambda)^{\tilde{\sigma}\sigma \frac{1-\mu}{\mu} } B_j(\lambda)^{\tilde{\sigma}(\sigma-1)\frac{1-\mu}{\mu}} L_j^{\tilde{\sigma}\gamma_2}, \\ \notag
\text{ with} \qquad  \gamma_1 &= 1 - (\sigma - 1) \alpha - \sigma \left(\frac{1-\mu}{\mu} \right) \tilde{\beta},\quad
	\gamma_2 = 1 + \sigma \alpha + (\sigma -1 ) \left( \frac{1-\mu}{\mu}\right) \tilde{\beta},
	\end{align*}
 and $\sigmatil$ defined as in the body of the article.

In our last step, we rewrite the above equation only in terms of the Voronoi weights. Using the expression for the weight and the welfare equalization condition for manufacturing workers, we find
\begin{equation*}
    L_i = \frac{\mu}{1-\mu} V^{-\frac{1}{\tilde\beta}} B_i(\lambda)^{-\frac{1}{\tilde{\beta}}} e^{-\frac{\delta}{\tilde{\beta}}\mu \lambda_i}
\end{equation*}
and substituting this expression into the equation above, we obtain
\begin{equation*}
	e^{-\frac{\delta \mu}{\tilde{\beta}} \tilde{\sigma}\gamma_1 \lambda_i } =  \left(\frac{1-\mu}{\mu} \right)^{(\sigma - 1)  \alpha} V^{\frac{\sigma -1}{\tilde \beta}(\alpha - \tilde \beta)} \ \sum_{j = 1}^{n^*} T_{ij}^{(1-\sigma)} \bar{A}_i^{\tilde{\sigma}(\sigma-1)} \bar{A}_j^{\tilde{\sigma}\sigma} B_i(\lambda)^{\sigmatil \phi_1} B_j(\lambda)^{\sigmatil \phi_2} e^{ -\frac{\delta \mu}{\tilde{\beta}} \tilde{\sigma}\gamma_2 \lambda_j },
	\end{equation*}
with $\phi_1$ and $\phi_2$ as in the main text.

\end{document}